\documentclass[final,5p,times,twocolumn]{elsarticle}

\usepackage{amssymb}
\usepackage{amsmath}
\usepackage{subcaption}
\usepackage{float} 
\usepackage{url}
\usepackage{adjustbox}
\usepackage{hyperref}

\hypersetup{
    colorlinks=true,
    linkcolor=blue,
    citecolor=blue,
    urlcolor=blue
}
\makeatletter
\def\ps@pprintTitle{%
    \let\@oddfoot\@empty
    \let\@evenfoot\@empty
    \def\@oddhead{%
        \hfil
        \footnotesize\itshape
        Published in \textit{Physics of the Dark Universe}
        \quad -- \quad
        DOI: \href{https://doi.org/10.1016/j.dark.2026.102464}
        {10.1016/j.dark.2026.102464}
        \hfil
    }%
    \let\@evenhead\@oddhead
}
\makeatother

\journal{Physics of the Dark Universe}

\begin{document}

\begin{frontmatter}


\title{Cosmography with gravitational waves I: Taylor expansion in the low-redshift regime}

%% Authors
\author[label1]{Jonathan Morais}
\ead{jmorais@on.br}

\author[label2,label1]{Rodrigo Gon\c{c}alves}
\ead{rsg\_goncalves@ufrrj.br}

\author[label1]{Jailson Alcaniz}
\ead{alcaniz@on.br}

%% Affiliations
\affiliation[label1]{organization={Observatório Nacional},
            city={Rio de Janeiro},
            postcode={20921-400},
            state={RJ},
            country={Brasil}}

\affiliation[label2]{organization={Departamento de Física, Universidade Federal Rural do Rio de Janeiro},
            city={Seropédica},
            postcode={23897-000},
            state={RJ},
            country={Brasil}}
            
%% Abstract
\begin{abstract}
We evaluate the capability of current and next-generation gravitational-wave detectors, including Advanced LIGO, the Einstein Telescope, and DECIGO, to probe cosmographic parameters using electromagnetically bright standard sirens. Adopting a third-order Taylor expansion, we examine the constraints primarily as a function of the number of detections and, in a complementary analysis, across correlated SNR--redshift regimes. Advanced LIGO can provide a distance-ladder-independent measurement of $H_0$ at the percent level. As the catalogue size increases, the Einstein Telescope reaches subpercent precision on $H_0$ and measures $q_0$ at the tens-of-percent level. DECIGO extends this sensitivity to $j_0$, achieving precision at the few-tens-of-percent level.
\end{abstract}

%% Keywords

\end{frontmatter}




\section{Introduction}\label{sec:introduction}

\indent The advent of multi-messenger gravitational wave (GW) astronomy has provided a novel research field to investigate both astrophysical processes and fundamental aspects of gravity and cosmology~\cite{2017Abbott}. The first GW signal~\cite{LIGOScientific:2016aoc} was detected on September 14, 2015, by the two detectors of the Laser Interferometer gravitational wave Observatory (LIGO)~\cite{LIGOScientific:2014pky}. Since then, the second-generation detectors, including LIGO, Virgo~\cite{VIRGO:2014yos}, and KAGRA~\cite{Somiya:2011np}, have confirmed more than 200 events~\cite{LIGOScientific:2025snk}, establishing GWs as a powerful new tool to investigate the Universe. In the next decade is expected the beginning of operations from the third-generation detectors such as the Einstein Telescope (ET)~\cite{Punturo:2010zz} and the Deci-hertz Interferometer gravitational wave Observatory (DECIGO)~\cite{Kawamura:2006up}. With their improved sensitivity, detections of coalescing binaries are expected to become as common as other probes, allowing us to study the Universe with much higher precision.\\
\indent From a cosmological perspective, however, even under optimistic assumptions, the current GW observations still have a limited impact. Among the hundreds of confirmed events nowadays, only one has a direct redshift measurement~\cite{LIGOScientific:2017vwq}. This measurement is crucial, since gravitational waves from compact binary coalescences can act as \textit{standard sirens}~\cite{Schutz:1986gp}: the GW analogs of standardizable candles~\cite{Branch:1992rv} in electromagnetic astronomy. This terminology arises because the characteristic \textit{chirp} waveform allows a direct measurement of the luminosity distance ($d_L$), independent of the cosmic distance ladder. However, the source redshift ($z$) cannot be uniquely determined from the GW signal alone due to a degeneracy between the intrinsic masses of the binary components and the redshift, which manifests as a degeneracy with $d_L$ and can be broken only with additional information.\\
\indent One such possibility arises when an electromagnetic counterpart is observed, enabling the identification of the host galaxy and, consequently, a direct measurement of the redshift. Several events involving at least one neutron star were followed up in search of such counterparts, including GW190425~\cite{LIGOScientific:2020aai}, as well as the neutron star–black hole mergers GW200105 and GW200115~\cite{LIGOScientific:2021qlt}; however, no confirmed association was found. This was successfully achieved for the binary neutron star merger GW170817, which was observed together with the short gamma-ray burst GRB170817A~\cite{LIGOScientific:2017zic}. Events of this kind, for which both the luminosity distance and the redshift are measured, are known as \textit{bright sirens}~\cite{Holz:2005df}. They provide a powerful and direct way to constrain cosmological parameters and to test the expansion history of the Universe~\cite{LIGOScientific:2017adf}. The next generation of interferometers is expected to detect a much larger number of bright sirens due to their enhanced sensitivities~\cite{Sathyaprakash:2009xs}, which will allow more precise and systematic studies of cosmology using gravitational wave observations~\cite{Zhang:2018byx, Wang:2021srv, Yang:2021qge}.\\
\indent Conversely, GW detections without an observed electromagnetic counterpart define the class of \textit{dark sirens}~\cite{Finke:2021aom}. In these cases, the redshift information must be inferred statistically, for instance through cross-correlation with galaxy catalogs~\cite{Sala:2025wwu}, probabilistic host-galaxy association or hierarchical Bayesian inference frameworks that combine population models with large-scale structure information~\cite{Mastrogiovanni:2023emh}. While dark sirens represent a promising way to extract cosmological information from the growing GW catalog, this work focuses on bright sirens, for which the redshift can be measured with negligible uncertainty. This enables a model-independent reconstruction of the luminosity distance–redshift relation $d_L(z)$, without assuming a specific cosmological background, within the framework of the so-called \textit{cosmographic approach}~\cite{Weinberg:1972kfs}.\\
\indent In this paper, we forecast the constraining power of current and next-generation gravitational-wave detectors, such as Advanced LIGO, the Einstein Telescope, and DECIGO, using electromagnetically bright standard sirens. While previous studies have explored standard sirens within a cosmographic framework~\cite{deSouza:2021xtg}, most works have primarily focused on constraining the Hubble constant $H_0$~\cite{Kyutoku:2016zxn, Chen:2017rfc, Chen:2024gdn, ET:2025xjr}. In contrast, we extend this approach by consistently implementing a higher-order cosmographic analysis and by providing a unified comparison across detectors with different sensitivities. Adopting a third-order Taylor expansion, we examine how the number of observed events  and the signal-to-noise ratio influence the reconstruction of the Hubble constant $H_0$, as well as the deceleration $q_0$ and jerk parameters $j_0$. The paper is organized as follows: Sec.~\ref{sec: cosmography} provides a concise overview of the cosmographic expansion. The simulated catalogs and the methodology adopted for the statistical analysis are described in Sec.~\ref{sec: Methodology}. The main results are presented and discussed in Sec.~\ref{sec: results}. Finally, Sec.~\ref{sec: conclusion} summarizes the main conclusions drawn from our analysis.

\section{Cosmography} \label{sec: cosmography}

\indent We begin by assuming a homogeneous, isotropic, and spatially flat Universe, described by the Friedmann–Lemaître–Robertson–Walker (FLRW) metric (with $c = 1$):
\begin{equation}
    ds^2 = -dt^2 + a^2(t) \left[dr^2 + r^2(d\theta^2 + \sin^2\theta\, d\phi^2)\right],
\end{equation}
where $t$ denotes cosmic time and $a(t)$ is the scale factor. In order to study the expansion history in a model-independent way, we follow~\cite{Visser:2004bf} and expand the scale factor as a Taylor series around the present cosmic time $t_0$:
\begin{align} \label{cosmography_a(t)}
\frac{a(t)}{a(t_0)} ={}& 1 + H_0 (t - t_0) - \frac{1}{2} q_0 [H_0 (t - t_0)]^2 
+ \frac{1}{6} j_0 [H_0 (t - t_0)]^3 \nonumber \\
\quad &+ \frac{1}{24} s_0 [H_0 (t - t_0)]^4 
+ \frac{1}{120} l_0 [H_0 (t - t_0)]^5 \nonumber \\
&+\frac{1}{720}m_0[H_0 (t - t_0)]^6 + \mathcal{O}(7),
\end{align}
where $H_0$ is the Hubble constant at the present epoch, and the higher-order derivatives of the scale factor define the following cosmographic parameters:
\begin{equation}
\label{cosmographic_parameters}
\begin{aligned}
q_0 &= -\frac{1}{H_0^2}
\left.\left(\frac{1}{a}\frac{d^2 a}{dt^2}\right)\right|_{t=t_0},
\qquad
j_0 &= \frac{1}{H_0^3}
\left.\left(\frac{1}{a}\frac{d^3 a}{dt^3}\right)\right|_{t=t_0},
\\[6pt]
s_0 &= \frac{1}{H_0^4}
\left.\left(\frac{1}{a}\frac{d^4 a}{dt^4}\right)\right|_{t=t_0},
\qquad
l_0 &= \frac{1}{H_0^5}
\left.\left(\frac{1}{a}\frac{d^5 a}{dt^5}\right)\right|_{t=t_0},
\\[6pt]
m_0 &= \frac{1}{H_0^6}
\left.\left(\frac{1}{a}\frac{d^6 a}{dt^6}\right)\right|_{t=t_0}.
\end{aligned}
\end{equation}

The parameters $q_0$, $j_0$, $s_0$, $l_0$ and $m_0$ are respectively referred to as the \textit{deceleration}, \textit{jerk}, \textit{snap}, \textit{lerk} and \textit{crackle} parameters. These quantities provide a purely kinematical description of the cosmic expansion and are defined by successive time derivatives of the scale factor evaluated at the present cosmic time.

By applying the definition of redshift, $1 + z \equiv a(t_0)/a(t)$ to Eq.~\eqref{cosmography_a(t)}, with $a(t_0) = 1$, one can express the Hubble parameter as a function of redshift~\cite{Zhang:2016urt}, given by:
\begin{align}
\label{H(z)}
H(z) &= H_0 \Bigg[\,
1 + (1+q_0)\,z
+ \frac{1}{2}\left(j_0-q_0^2\right)z^2
\nonumber \\[4pt]
&\quad
+ \frac{1}{6}\Bigl(
3q_0^2 + 3q_0^3
-4q_0j_0 -3j_0 -s_0
\Bigr)z^3
\nonumber \\[4pt]
&\quad
+ \frac{1}{24}\Bigl(
-12q_0^2 -24q_0^3 -15q_0^4
+32q_0j_0 +25q_0^2j_0
\nonumber \\[4pt]
&\qquad
+7q_0s_0 +12j_0 -4j_0^2
+8s_0 +l_0
\Bigr)z^4
\nonumber \\[4pt]
&\quad
+ \frac{1}{120}\Bigl(
m_0
-60q_0^2
-180q_0^3
-225q_0^4
-105q_0^5
\nonumber \\[4pt]
&\qquad
-10j_0^2(6+7q_0)
+l_0(15+11q_0)
\nonumber \\[4pt]
&\qquad
+15j_0\bigl(
4+16q_0+25q_0^2+14q_0^3-s_0
\bigr)
\nonumber \\[4pt]
&\qquad
+60s_0
+105q_0s_0
+60q_0^2s_0
\Bigr)z^5
+ \mathcal{O}(z^6)
\,\Bigg].
\end{align}

We can use Eq.~\eqref{H(z)} to define other distance functions relevant in cosmology. In particular, for our analysis it is convenient to express the \textit{luminosity distance} in terms of the cosmographic expansion
\begin{align} \label{dl_cosmography}
d_L(z) = &\frac{1}{H_0} \Bigg[  \nonumber
z 
+ \frac{1}{2}(1 - q_0) z^2 
+ \left(-\frac{1}{6} - \frac{j_0}{6} + \frac{q_0}{6} + \frac{q_0^2}{2}\right) z^3 \\ \nonumber
&+ \left(\frac{1}{12} + \frac{5 j_0}{24} - \frac{q_0}{12} + \frac{5 j_0 q_0}{12} 
- \frac{5 q_0^2}{8} - \frac{5 q_0^3}{8} + \frac{s_0}{24}\right) z^4 \\ \nonumber
&+ \Bigg(-\frac{1}{20} - \frac{9 j_0}{40} + \frac{j_0^2}{12} - \frac{l_0}{120} + \frac{q_0}{20}
- \frac{11 j_0 q_0}{12}\\ \nonumber 
&+ \frac{27 q_0^2}{40} - \frac{7 j_0 q_0^2}{8} 
+ \frac{11 q_0^3}{8} + \frac{7 q_0^4}{8} - \frac{11 s_0}{120} - \frac{q_0 s_0}{8}\Bigg) z^5 \\ \nonumber
&+ \Bigg(\frac{1}{30} + \frac{7 j_0}{30} - \frac{19 j_0^2}{72} + \frac{19 l_0}{720} + \frac{m_0}{720}
- \frac{q_0}{30} + \frac{13 j_0 q_0}{9}\\ \nonumber 
&- \frac{7 j_0^2 q_0}{18} + \frac{7 l_0 q_0}{240} - \frac{7 q_0^2}{10} + \frac{133 j_0 q_0^2}{48}
- \frac{13 q_0^3}{6} + \frac{7 j_0 q_0^3}{4}\\ 
&- \frac{133 q_0^4}{48} - \frac{21 q_0^5}{16} + \frac{13 s_0}{90} - \frac{7 j_0 s_0}{144}
+ \frac{19 q_0 s_0}{48}\\ \nonumber
&+ \frac{7 q_0^2 s_0}{24}\Bigg) z^6 + \mathcal{O}(z^7)
\Bigg].
\end{align}

The Taylor expansion employed in cosmography suffers from convergence issues at high redshifts. As discussed by Visser~\cite{Visser:2004bf}, the series expansion of cosmological quantities in terms of the redshift $z$ has a finite radius of convergence, set by the nearest singularity in the complex $z$-plane. In particular, the transformation $1+z = a(t_0)/a(t)$ implies the presence of a pole at $z=-1$, corresponding to the limit $a \to \infty$. As a consequence, the Taylor series about $z=0$ converges only within $|z|<1$. Therefore, although the series is formally defined for positive redshifts, it becomes unreliable for $z \gtrsim 1$, where the expansion lies outside its radius of convergence and truncation errors grow rapidly. This limitation directly affects the luminosity distance $d_L(z)$, whose Taylor series can diverge or lose precision for sources at intermediate or high redshifts, thus compromising  its reliability in those regimes. To achieve better convergence at larger redshifts, alternative parameterizations have been proposed, such as the $y$-redshift~\cite{Cattoen:2007sk,Lazkoz:2013ija}, or rational approximations like the Padé~\cite{Gruber:2013wua}, Chebyshev~\cite{Shafieloo:2012jb} expansions, among others~\cite{Hu:2022udt}.

However, exploring these alternative formulations lies beyond the scope of the present work, where we restrict our analysis to the standard expansion in terms of the usual $z$-redshift. Therefore, to remain within the convergence domain, we restrict our analysis to data with redshifts $z<1$. Within this regime, we quantify the truncation error at successive orders and determine the redshift interval over which the third-order approximation remains sufficiently accurate. This analysis is presented in detail in the following section.

\section{Methodology} \label{sec: Methodology}

\begin{figure*}[!t]
    \centering

    \begin{subfigure}{0.48\textwidth}
        \centering
        \includegraphics[
            width=\linewidth,
            keepaspectratio
        ]{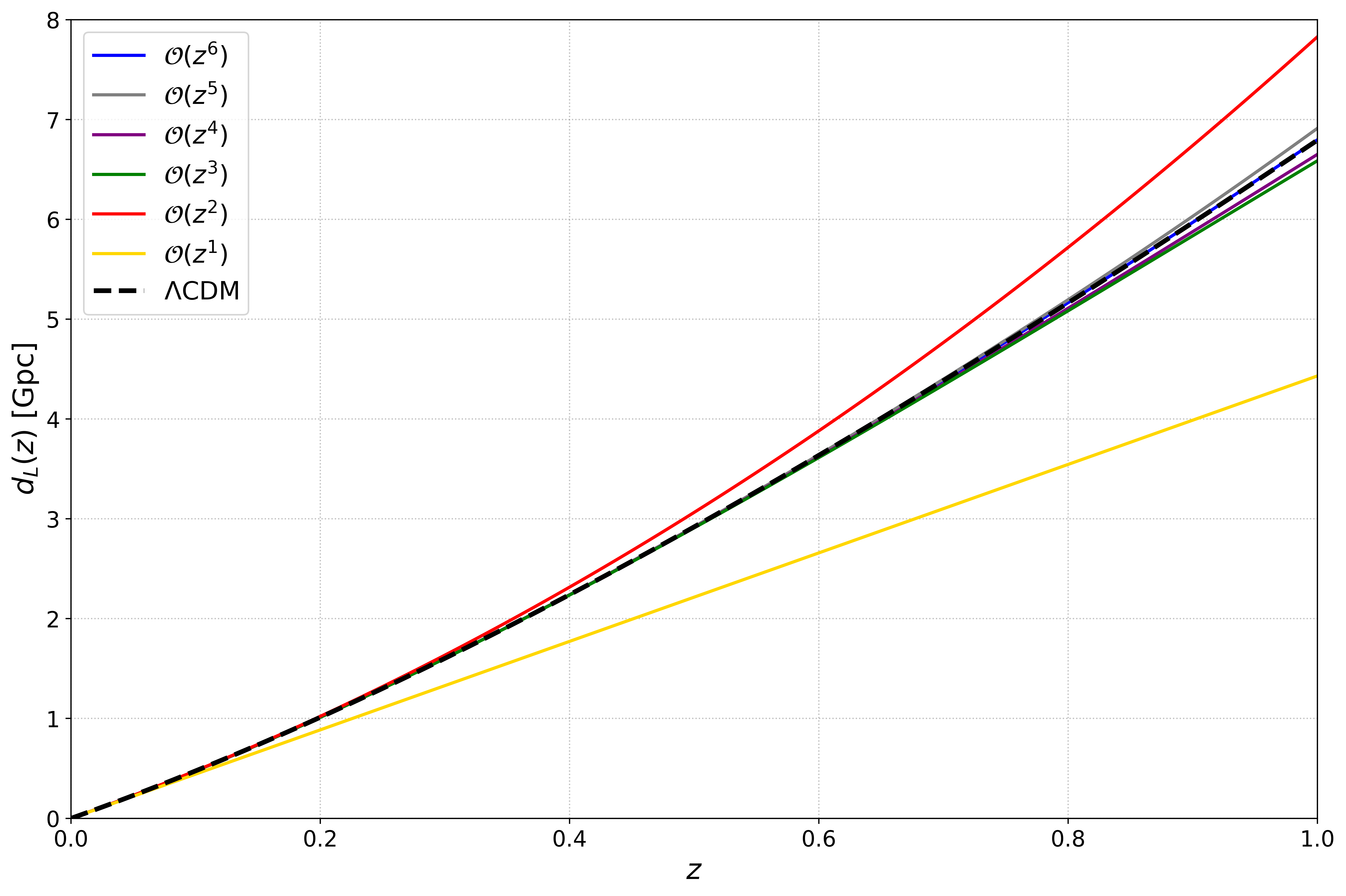}
        \label{fig:dl_differents_orders}
    \end{subfigure}
    \hfill
    \begin{subfigure}{0.48\textwidth}
        \centering
        \includegraphics[
            width=\linewidth,
            keepaspectratio
        ]{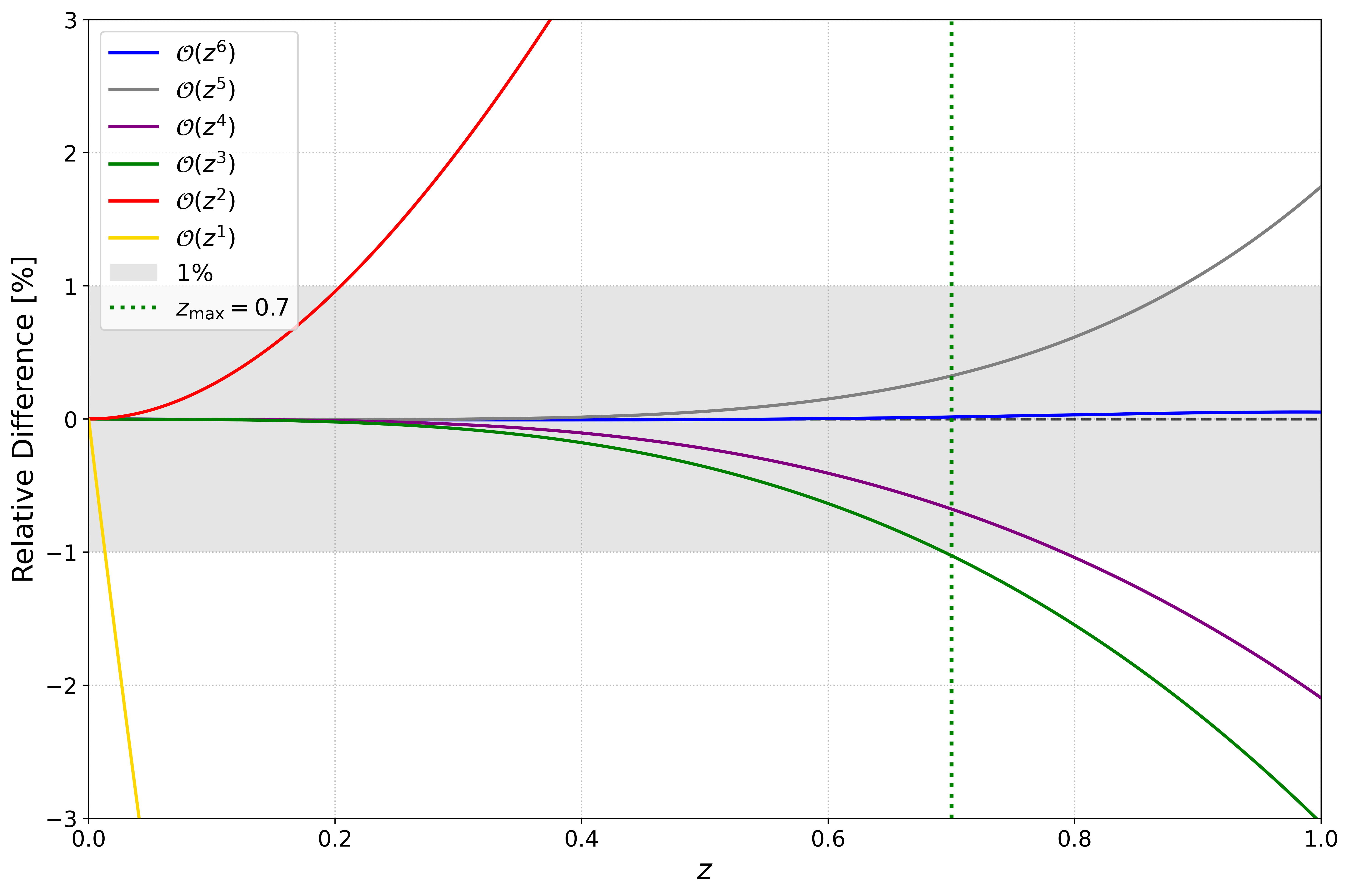}
        \label{fig:relative_difference_orders}
    \end{subfigure}

    \caption{
    Luminosity distance as a function of redshift for different orders
    of the cosmographic series (left panel).
    The black dashed curve corresponds to the fiducial $\Lambda$CDM
    model used in the analysis.
    Relative difference between each cosmographic series and the
    fiducial model (right panel).
    The gray shaded region indicates a $1\%$ deviation, which is
    reached by the third-order series at $z \approx 0.7$
    (vertical dotted line).
    }

    \label{figure_1}
\end{figure*}

In order to investigate how the luminosity distance expansion at different orders deviates from a given cosmological model, we first define the fiducial cosmology that will serve as the reference framework for our simulations. This model corresponds to the standard $\Lambda$CDM cosmology, characterized by the matter density parameter $\Omega_m = 0.311$ and the Hubble constant $H_0 = 67.66$ km\,s$^{-1}$\,Mpc$^{-1}$, in agreement with the most recent Cosmic Microwave Background (CMB) measurements~\cite{Planck:2018vyg}. It is worth noting that the same theoretical analysis was also performed using the SH0ES estimations for both parameters~\cite{Riess:2021jrx}, yielding practically identical results. Therefore, the particular choice of $H_0$ does not significantly affect our conclusions.\\
\indent To determine the values of the cosmographic parameters, we adopt the analytical expressions derived within the $\Lambda$CDM framework~\cite{Pourojaghi:2024bxa}. Explicitly, the cosmographic parameters in Eq.~\eqref{cosmographic_parameters} can be expressed in terms of one another (see Eq.~(7) in Ref.~\cite{Pourojaghi:2024bxa}). By substituting the standard $\Lambda$CDM Hubble function $H(z) = H_0 \sqrt{\Omega_m (1+z)^3 + 1 - \Omega_m}$, we obtain the corresponding relations 
\begin{align} 
q_0 &= \frac{3}{2}\Omega_m - 1, \nonumber \\
j_0 &= 1, \nonumber \\
s_0 &= 1 - \frac{9}{2}\Omega_m, \nonumber \\
l_0 &= 1 + 3\Omega_m + \frac{27}{2}\Omega_m^2, \nonumber \\
m_0 &= 1 - \frac{27}{2}\Omega_m - 81\Omega_m^2
       - \frac{81}{4}\Omega_m^3.
\label{LCDM_cosmography}
\end{align}
Using the abovementioned matter density parameter in combination with Eq.~\eqref{LCDM_cosmography}, the numerical values of the corresponding cosmographic parameters are summarized in Table~\ref{tab_fiducial_parameters}.
\begin{table}[h!]
\centering
\begin{tabular}{cccccc}
\hline
$q_0$ & $j_0$ & $s_0$ & $l_0$ & $m_0$ \\
\hline
$-0.533$ & $1$ & $-0.400$ & $3.239$ & $-11.642$ \\
\hline
\end{tabular}
\caption{Cosmographic parameters in a flat $\Lambda$CDM cosmology with $\Omega_m = 0.311$.}
\label{tab_fiducial_parameters}
\end{table}
We can now calculate the luminosity distance using Eq.~\eqref{dl_cosmography} and the fiducial parameters listed in Table~\ref{tab_fiducial_parameters}, considering the redshift range $z < 1$ due to the convergence issues discussed previously. In Fig.~\ref{figure_1}, we present the behavior of $d_L$ as a function of redshift for different truncation orders of the cosmographic series. For comparison, the fiducial $\Lambda$CDM model is also shown (black dashed curve). The bottom panel displays the relative difference between each cosmographic approximation and the fiducial model.
Within the convergence domain, truncating the Taylor series at progressively higher orders is expected to provide a more accurate representation of the luminosity distance. This behavior is illustrated in the top panel of Fig.~\ref{figure_1}, where the successive approximations increasingly approach the fiducial $\Lambda$CDM curve. However, each additional term introduces a new cosmographic parameter, which may increase degeneracies and statistical uncertainties. We therefore choose to truncate the series at third order, restricting the analysis to $H_0$, $q_0$, and $j_0$. This conservative choice is supported by observational studies based specifically on Taylor cosmography. Using the standard expansion in $z$ and the JLA sample, Zhang et al.~\cite{Zhang:2016urt} found that terminating the series at the jerk term provides the best balance between accuracy and statistical variance. Similarly, Hu and Wang~\cite{Hu:2022udt} reported that the Pantheon data slightly favor the third-order form over its fourth-order counterpart according to the AIC and BIC criteria.\\
\indent Having fixed the truncation order, we determine its range of validity by requiring a maximum relative deviation of $1\%$ from the fiducial $\Lambda$CDM prediction. This tolerance keeps the error associated with the neglected terms at the percent level and provides an objective criterion for defining the upper redshift limit. The bottom panel of Fig.~\ref{figure_1} shows that the third-order result reaches the adopted threshold at $z\approx0.7$, while all subsequent orders remain within the same limit over this interval. The corresponding redshift for each case is reported in Table~\ref{tab:zmax_values}. We therefore adopt $z_{\mathrm{max}}=0.7$ for our dataset.
\begin{table}[ht]
    \centering
    \begin{tabular}{cc}
        \hline
        {Order} & {$z_{\text{max}}$ for $1\%$ deviation} \\
        \hline
        $\mathcal{O}(z^1)$ & 0.0133 \\
        $\mathcal{O}(z^2)$ & 0.2047 \\
        $\mathcal{O}(z^3)$ & 0.6942 \\
        $\mathcal{O}(z^4)$ & 0.7901 \\
        $\mathcal{O}(z^5)$ & 0.8868 \\
        $\mathcal{O}(z^6)$ & $>1$ \\
        \hline
    \end{tabular}
    \caption{Redshift values at which each cosmographic expansion order reaches a maximum deviation of $1\%$ from the fiducial $\Lambda$CDM model. For $\mathcal{O}(z^6)$, this threshold is not reached within the adopted convergence domain, $z<1$.}
    \label{tab:zmax_values}
\end{table}

\subsection{Gravitational Waves Source Distribution} \label{sec: GW_distribution}

\begin{figure}[htbp]
    \centering
    \includegraphics[width=0.45\textwidth]{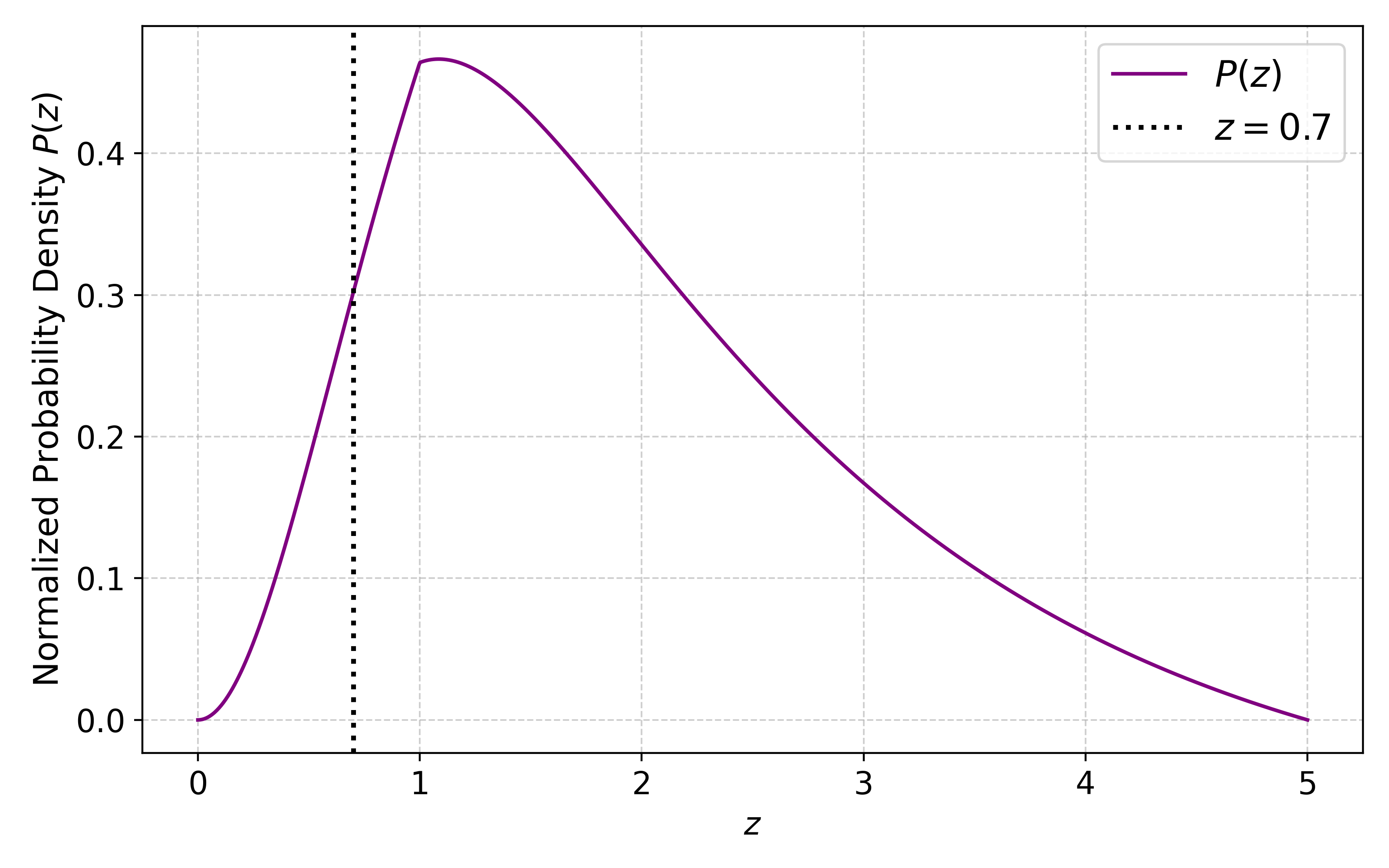}
    \caption{Normalized redshift probability density function, $P(z)$, for stellar-mass binary systems. The vertical dashed line marks the maximum redshift considered in the analysis, $z_{\rm max} = 0.7$.}
    \label{fig:redshift_distributions}
\end{figure}

The redshift distribution of coalescing sources provides a statistical description of their spatial distribution. For stellar-origin binaries, the redshift probability distribution~\cite{Zhao:2010sz} can be expressed as
\begin{equation} \label{P(z)_stelar_binaries}
    P(z) \propto \frac{4\pi d^{2}_{C}(z)R(z)}{H(z)(1+z)},
\end{equation}
where $d_C(z) = \int^z_0 H^{-1}(z')dz'$ is the comoving distance, and $R(z)$ describes the evolution of the merger rate following the cosmic star formation history (see~\cite{Schneider:2000sg}). Even though the redshift reach of gravitational wave sources can extend up to $z \approx 5$ for stellar-origin binaries, we adopt a detection limit of $z_{\text{max}} = 0.7$ in our analysis to ensure the convergence of the cosmographic series, as explained before. The redshift distribution, given by Eq.~\eqref{P(z)_stelar_binaries} with $\Lambda$CDM as a fiducial cosmology, is shown in  Fig.~\ref{fig:redshift_distributions}.

In this work, we simulate gravitational wave data from third generation detectors, namely the Einstein Telescope and DECIGO, and the Advanced LIGO (aLIGO) is also included as a reference for current ground-based sensitivity. The interferometers considered in this work span complementary frequency ranges, from the decihertz band targeted by DECIGO to the higher-frequency bands probed by the ground-based aLIGO and ET detectors. Within these frequency ranges, binary neutron star (BNS) mergers are expected to be among the dominant bright-siren sources~\cite{Patricelli:2022hhr, Liu:2022mcd}.\\
\indent The expected number of gravitational wave detections ($N_e$) strongly depends on the detector sensitivity, design, and mission duration. For the Einstein Telescope, approximately $10^3$ bright sirens are expected over five years of operation, with sources detectable up to $z \approx 3$~\cite{ETDesignReport2020}. DECIGO, due to its higher sensitivity, is predicted to observe around $10^4$ bright sirens, reaching $z \approx 5$~\cite{Liu:2024gne}. In contrast, there is no clear consensus in the literature regarding the expected number of bright siren detections for Advanced LIGO, which is limited to $z \lesssim 0.2$~\cite{Lagos:2019kds}. Therefore, we adopt a conservative estimate of one event per year~\cite{Schutz:2018fso}, resulting in 10 detections over a decade of operation.\\
\indent It is important to note that the detection rates quoted for each instrument refer to their full redshift range. Since our analysis is restricted to simulations up to $z_{\text{max}} = 0.7$, the corresponding number of detectable events is lower within our redshift interval of interest.\\
\indent To estimate this effective number of detections, we first generate random realizations of source redshifts following the probability density functions defined in Eq.~\eqref{P(z)_stelar_binaries} over the full redshift range. We then restrict the sample to only those points within the redshift range $0 < z < 0.7$ for each detector. By performing this analysis multiple times we average the number of detections simulated and define them as $N_z$, which are summarized in Table~\ref{tab:N_initial}.
    
\begin{table}[h]
\centering
\caption{Expected number of detections ($N_e$) for each interferometer and the adopted number of events ($N_z$) within the redshift limit $z_{\text{max}} = 0.7$.}
\label{tab:N_initial}
\begin{tabular}{lcc}
\hline
\textbf{Interferometer} & $\boldsymbol{N_e}$ & $\boldsymbol{N_z}$ \\
\hline
Advanced LIGO & 10    & 10  \\
Einstein Telescope  & 1000  & 100 \\
DECIGO & 10000 & 800 \\
\hline
\end{tabular}
\end{table}

\subsection{Gravitational Waves Detection} \label{GW_detection}

With all the relevant information regarding the gravitational wave sources established, the next step is to simulate how these signals would be detected by ground and space-based interferometers. In the transverse-traceless (TT) gauge, the response of a detector to an incoming GW can be expressed as
\begin{equation}
    h(t) = F_{+}(\theta,\phi,\psi)\,h_{+}(t) + F_{\times}(\theta,\phi,\psi)\,h_{\times}(t),
\end{equation}
where $F_{+}$ and $F_{\times}$ represent the antenna pattern functions corresponding to the two independent gravitational wave polarizations~\cite{Sathyaprakash:2009xs}. The explicit analytical forms of these functions for the aLIGO, Einstein Telescope and DECIGO detectors are presented respectively in~\cite{Yang:2021qge, Califano:2022cmo, Yagi:2011wg}. The polarization components are defined as $h_{+} = h_{xx} = -h_{yy}$ and $h_{\times} = h_{xy} = h_{yx}$. 

The angular parameters $(\theta, \phi)$ describe the sky position of the source in the detector’s reference frame, while $\psi$ denotes the polarization angle. The specific choices adopted for these angular variables in the simulations of each detector follow the prescription described in~\cite{Cai:2017aea}. It is worth emphasizing that the antenna pattern functions depend sensitively on the geometric configuration of the interferometer, particularly on the relative orientation and the opening angle between its arms, which ultimately determine the detector’s directional sensitivity.

To analyze the signal in the frequency domain, we apply the stationary phase approximation, which yields
\begin{equation} \label{phase_eq}
    \mathcal{H}(f) = \mathcal{A}f^{-7/6}e^{i\Psi(f)},
\end{equation}
where the amplitude $\mathcal{A}$ is given by
\begin{equation} \label{amplitude}
\mathcal{A} = \frac{1}{d_L} 
\sqrt{F_+^2 (1 + \cos^2\iota)^2 + 4 F_\times^2 \cos^2\iota}
\, \sqrt{\frac{5\pi}{96\pi}} \, \mathcal{M}_c^{5/6} \, \pi^{-7/6}.
\end{equation}
Here, $\iota$ denotes the inclination angle between the orbital plane of the binary system and the observer’s line of sight. The quantity $\mathcal{M}_c$ corresponds to the observed chirp mass, defined as $\mathcal{M}_c = (1+z)M\eta^{3/5}$, where $M = m_1 + m_2$ is the total mass of the system and $\eta = \frac{m_1 m_2}{M^2}$ represents the symmetric mass ratio. 

In this work, we assume uniform mass distributions for the compact objects composing the binaries~\cite{Cai:2017aea}. For stellar-origin systems, the neutron star masses are uniformly distributed within the interval $(1,2)\,M_\odot$, where $M_\odot$ denotes the solar mass.

The phase $\Psi(f)$ in Eq.~\eqref{phase_eq} is evaluated within the post-Newtonian framework up to 3.5PN order~\cite{Nishizawa_2011}. Within this approximation, spin effects are neglected, as their contribution to the waveform phase is subdominant for non-spinning or slowly spinning systems.

Once the amplitude and phase of the waveform are determined, the next step is to assess whether the detected strain corresponds to an actual GW signal or is merely a product of instrumental noise. A detection is claimed only when the signal-to-noise ratio (SNR) of the detector network exceeds a threshold of 8, consistent with the criterion currently adopted by the LIGO and Virgo collaborations~\cite{KAGRA:2013rdx}.

\begin{figure*}[!t]
    \centering

    \begin{subfigure}{0.48\textwidth}
        \centering
        \includegraphics[width=\linewidth]{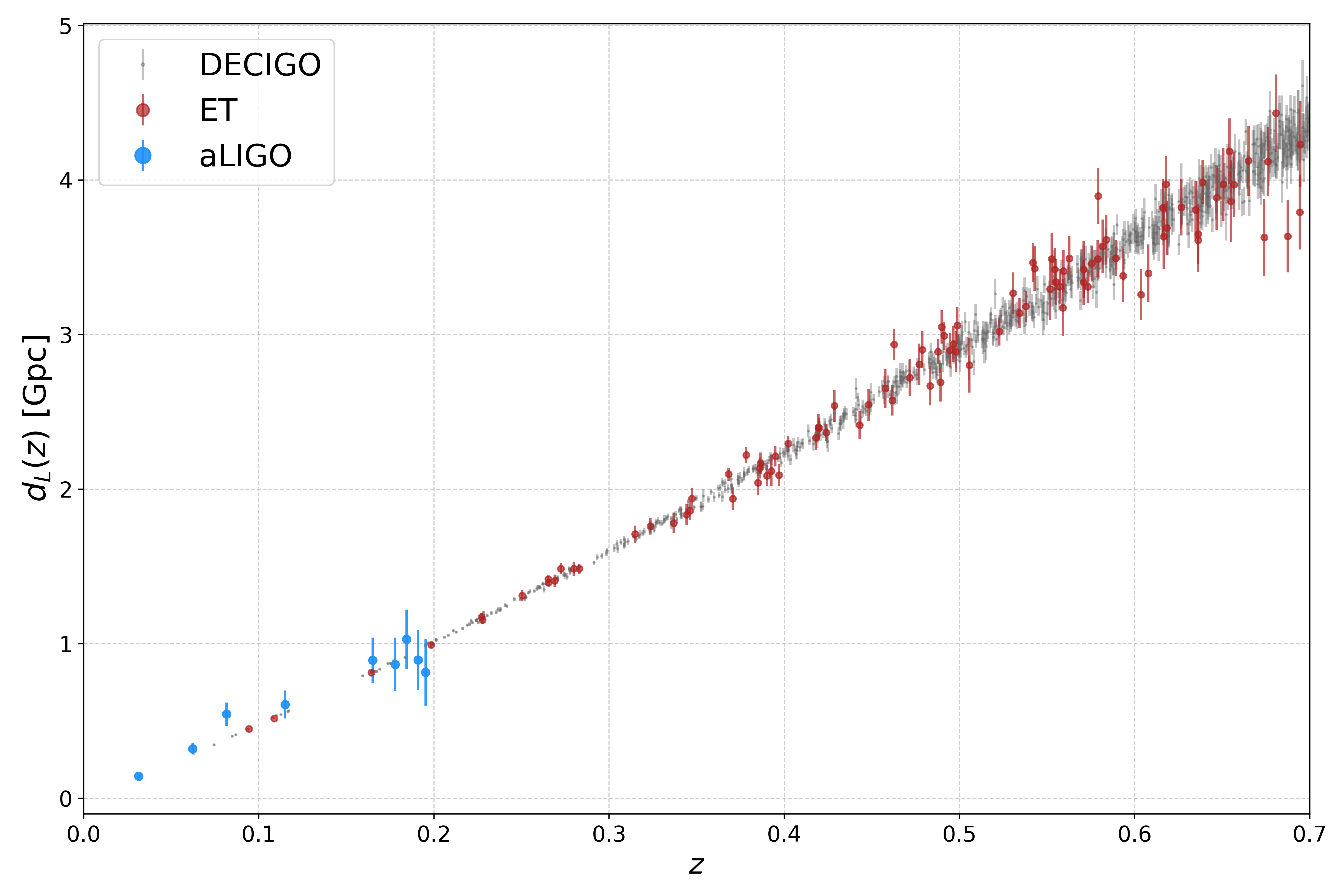}
        \label{fig:dl_sim}
    \end{subfigure}
    \hfill
    \begin{subfigure}{0.48\textwidth}
        \centering
        \includegraphics[width=\linewidth]{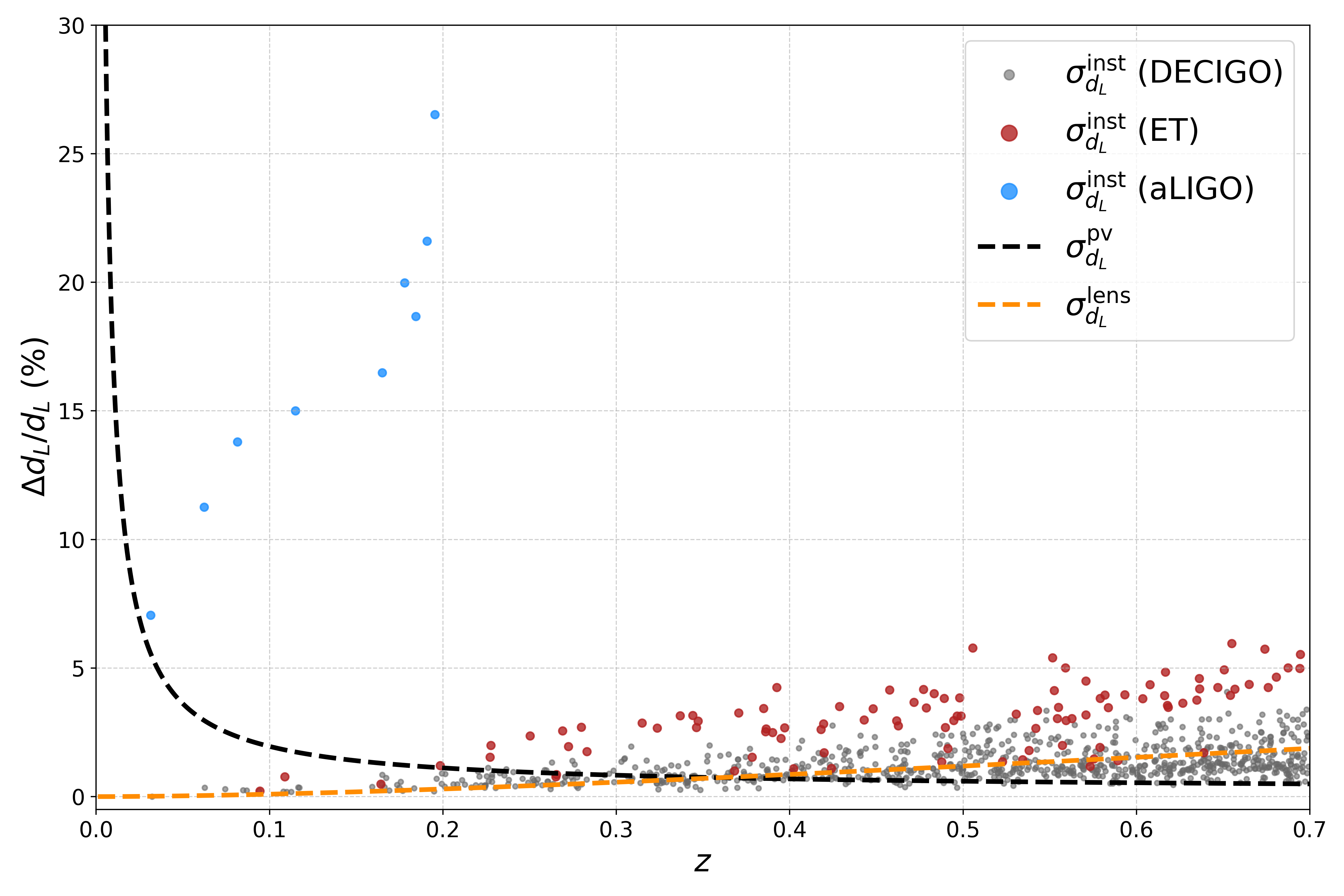}
        \label{fig:sigma_frac}
    \end{subfigure}

    \caption{
    Simulated catalogs of luminosity distance for each interferometer
    (DECIGO, ET and aLIGO), with the number of detections set to $N_z$,
    a maximum redshift of $z_{\mathrm{max}} \simeq 0.7$ and
    $\Lambda$CDM as the fiducial cosmology (left panel).
    Fractional uncertainties on the luminosity distance, where the points
    correspond to the instrumental errors extracted from the simulated
    catalogs, and the dashed and solid curves represent the systematic
    contributions from weak lensing and peculiar velocities, respectively,
    expressed as $\Delta d_L / d_L \times 100\%$ (right panel).
    }

    \label{fig:simulations_erros}
\end{figure*}

For a network composed of $N$ independent interferometers, the combined SNR is defined as
\begin{equation} \label{SNR}
    \rho = \sqrt{\sum_{i=1}^{N} \left( \rho^{(i)} \right)^2},
\end{equation}
where $\rho^{(i)} = \sqrt{\langle \mathcal{H}^{(i)}, \mathcal{H}^{(i)} \rangle}$ denotes the SNR of the $i$-th detector.

The inner product used in Eq.~\eqref{SNR} is computed as
\begin{equation}
\langle a, b \rangle \equiv 4 \int_{f_{\text{min}}}^{f_{\text{max}}} 
\frac{\tilde{a}(f)\tilde{b}^*(f) + \tilde{a}^*(f)\tilde{b}(f)}{2} 
\frac{df}{S_h(f)},
\end{equation}
where $\tilde{a}(f)$ and $\tilde{b}(f)$ are the Fourier transforms of the time-domain functions $a(t)$ and $b(t)$, and $S_h(f)$ is the one-sided noise power spectral density (PSD) of the detector. The frequency limits $f_{\min}$ and $f_{\max}$ depend on the specific interferometer configuration~\cite{Cai:2017aea, Yang:2017gxh}. The PSD functions for Advanced LIGO and the Einstein Telescope are taken from the official technical documentation of the respective instruments\footnote{\url{https://dcc.ligo.org/LIGO-T2000012/public}}\footnote{\url{https://www.et-gw.eu/index.php/etsensitivities}}. Additional details on the aLIGO design sensitivity are provided in Ref.~\cite{KAGRA:2013rdx}. For the ET, we adopt the ET-D sensitivity configuration~\cite{Hild:2010id}. The DECIGO PSD is implemented following the description given in Ref.~\cite{Liu:2024gne}.

After confirming the presence of a signal via Eq.~\eqref{SNR}, we proceed to estimate the measurement uncertainty in the luminosity distance. The instrumental contribution to this uncertainty is evaluated within the Fisher matrix formalism~\cite{Cai:2017aea}. As a first step, we consider a simplified estimate in which the uncertainty in the luminosity distance $d_L$ is treated as uncorrelated with the remaining gravitational-wave parameters. Under this assumption, the instrumental error can be written as
\begin{equation}
    \sigma^{\text{inst}}_{d_L} \simeq \sqrt{\left\langle \frac{\partial H}{\partial d_L}, \frac{\partial H}{\partial d_L} \right\rangle^{-1}}.
\end{equation}
From Eqs.~\eqref{phase_eq} and \eqref{amplitude}, the waveform amplitude scales as $H \propto d_L^{-1}$, leading to the well-known relation $\sigma^{\text{inst}}_{d_L} \simeq d_L / \rho$ in the absence of parameter correlations. This estimate is subsequently refined by accounting for the dominant correlation between $d_L$ and the inclination angle $\iota$. The inclination can modify the signal-to-noise ratio by up to a factor of two for $\iota$ in the range $0^\circ$–$90^\circ$~\cite{Cai:2017aea}. To conservatively incorporate this effect, we rescale the uncertainty, obtaining
\begin{equation} \label{inst_error}
    \sigma^{\text{inst}}_{d_L} \simeq \frac{2d_L}{\rho}.
\end{equation}

An additional contribution to the luminosity distance uncertainty arises from weak gravitational lensing ($\sigma^{\text{lens}}_{d_L}$)~\cite{Zhao:2010sz} which can be expressed as
\begin{equation}
    \sigma_{d_L}^{\text{lens}}  =   0.05\, z\, d_L.
\end{equation}

We also account for the uncertainty due to peculiar velocities ($\sigma^{\text{pv}}_{d_L}$) arising from galaxy clustering and the motion of the binary barycenter~\cite{Gordon:2007zw}. This contribution is modeled as
\begin{equation}
    \sigma^{\text{pv}}_{d_L}(z) = d_L(z) \times 
    \left| 1 - \frac{(1+z)^2}{H(z) d_L(z)} \right| 
    \sigma_{v, \text{gal}},
\end{equation}
where $\sigma_{v, \text{gal}}$ denotes the one-dimensional velocity dispersion of galaxies, typically taken as $\sigma_{v, \text{gal}} = 300\, \mathrm{km\,s^{-1}}$, independent of redshift. The total uncertainty in the luminosity distance is then obtained by combining these contributions in quadrature:
\begin{equation}
    \sigma_{d_L} = \sqrt{(\sigma^{\text{inst}}_{d_L})^2 + (\sigma^{\text{lens}}_{d_L})^2 + (\sigma^{\text{pv}}_{d_L})^2}.
\end{equation}

Once $d_L$ and $\sigma_{d_L}$ are obtained, one may perform a Monte Carlo simulation to generate mock GW catalogues by sampling luminosity distances from a Gaussian distribution centered on the fiducial $d_L$ values.

In Figure~\ref{fig:simulations_erros}, we present a summary of the simulated data used in our analysis. The top panel shows one realization of the luminosity distance for each detector (aLIGO, ET and DECIGO), constructed following the full procedure described in this section. The bottom panel displays the fractional instrumental errors, together with the systematic contributions arising from weak lensing and peculiar velocities. As can be observed, the fractional error associated with peculiar velocities dominates at low redshifts, since small variations in local motion correspond to a significant fraction of the total luminosity distance in this regime~\cite{Blake:2025etn}. Conversely, the lensing contribution increases with the redshift, reflecting the cumulative distortion of light paths caused by the large-scale structure of the Universe~\cite{Canevarolo:2023dkh}. The behavior of the instrumental errors depends on the characteristics of each detector, particularly on its SNR. As shown, the next-generation interferometers exhibit significantly smaller instrumental uncertainties than the current aLIGO configuration, mainly due to their enhanced sensitivity.

\begin{figure*}[t!]
\centering

\begin{subfigure}{\textwidth}
\centering
\includegraphics[width=0.65\textwidth]{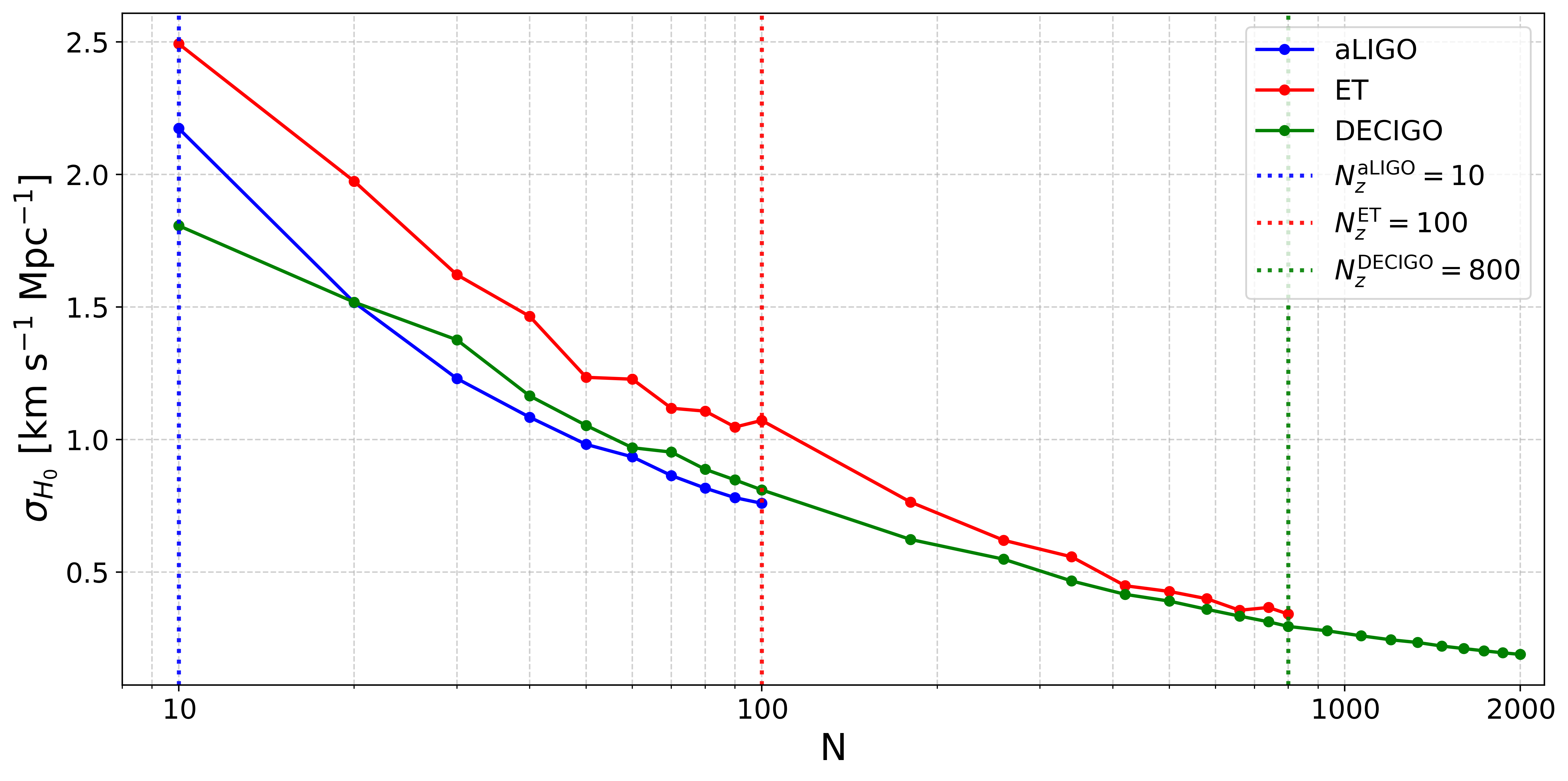}
%\caption{}
\label{fig:sigma_H0_N}
\end{subfigure}

%\vspace{1cm}

\begin{subfigure}{\textwidth}
\centering
\includegraphics[width=0.65\textwidth]{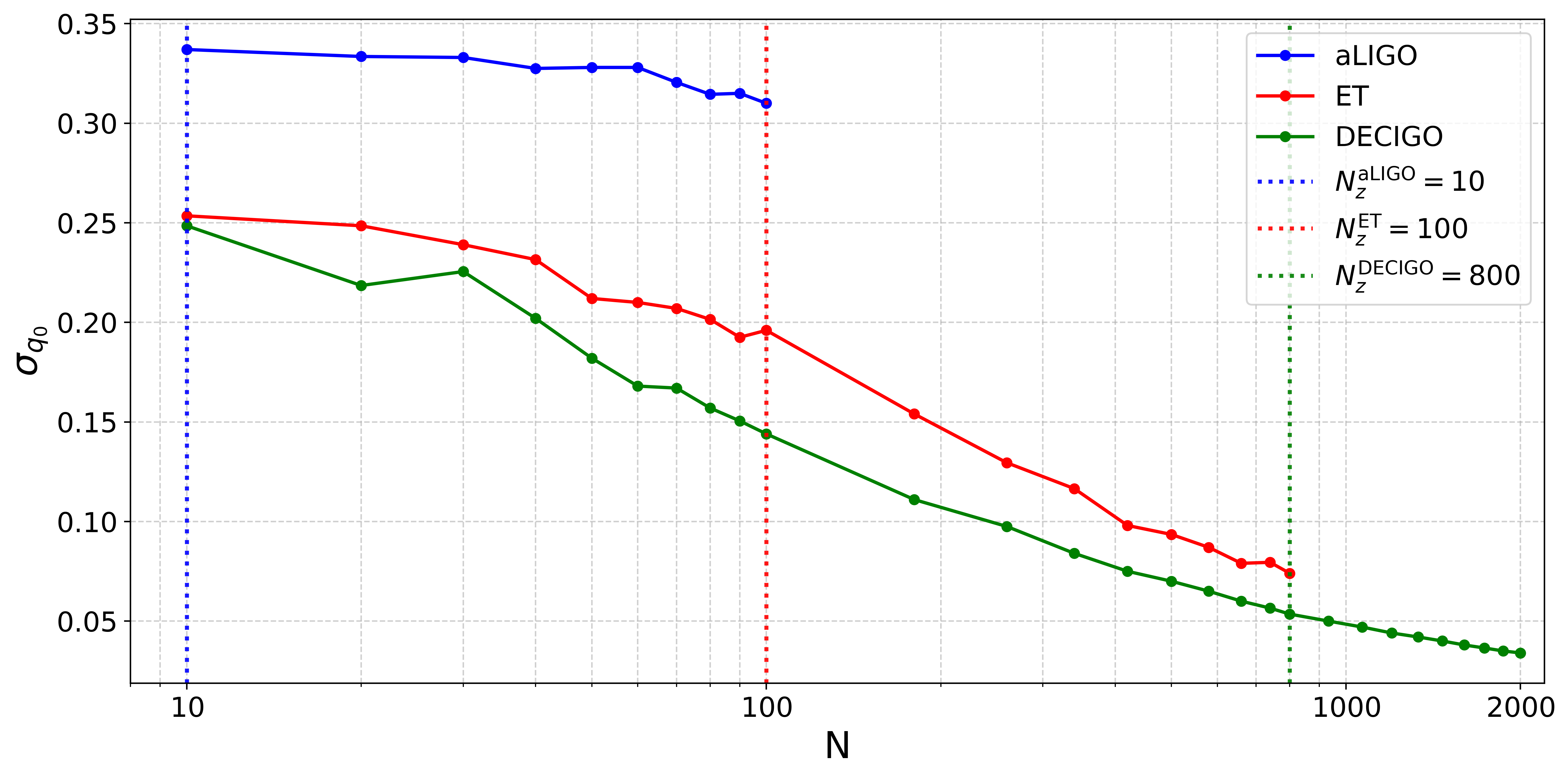}
%\caption{}
\label{fig:sigma_q0_N}
\end{subfigure}

%\vspace{1cm}

\begin{subfigure}{\textwidth}
\centering
\includegraphics[width=0.65\textwidth]{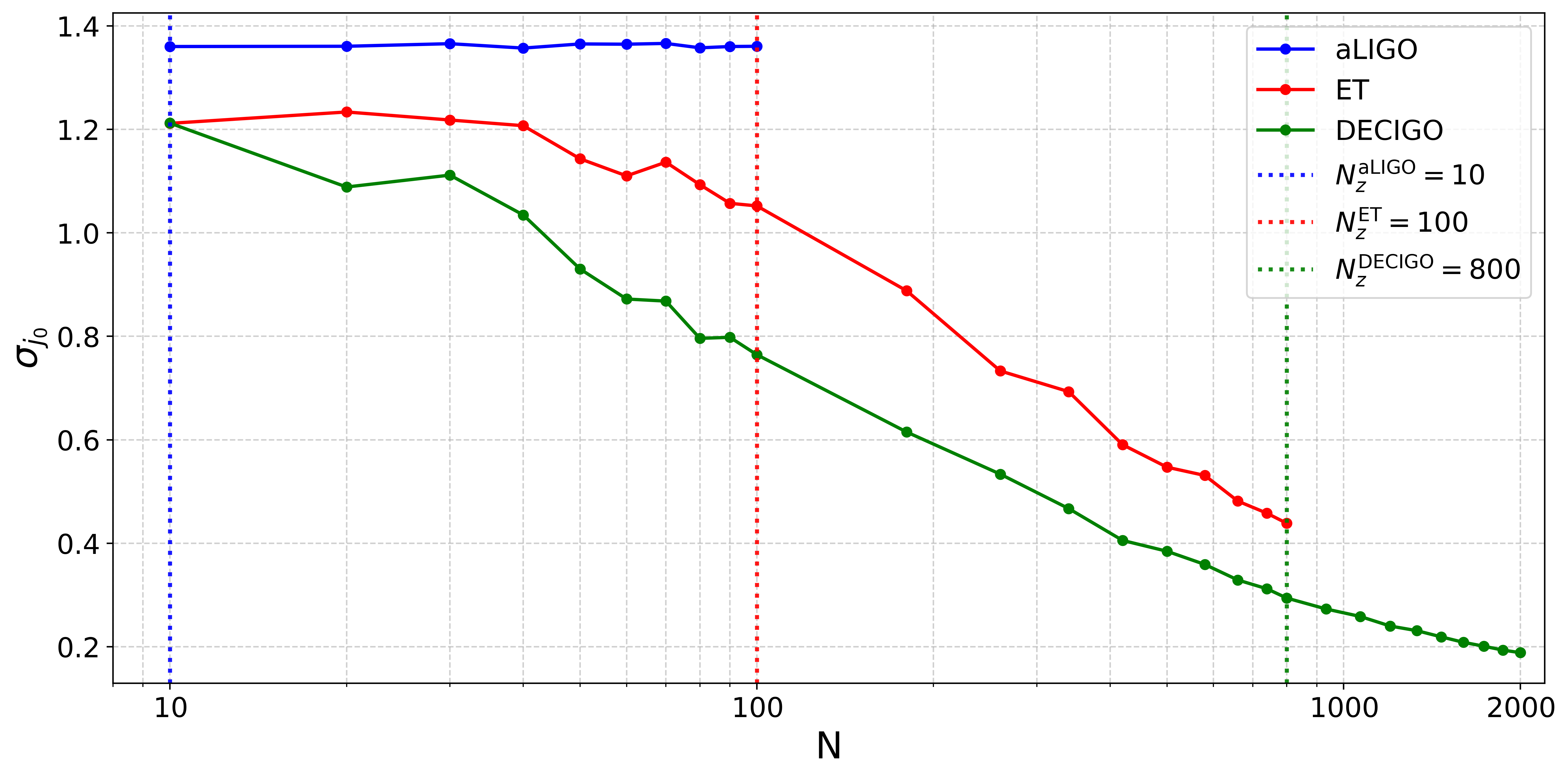}
%\caption{}
\label{fig:sigma_j0_N}
\end{subfigure}

\caption{Uncertainties, $\sigma$, of the cosmographic parameters $H_0$, $q_0$, and $j_0$ as functions of the number of detections, $N$. The top, middle, and bottom panels correspond to $H_0$, $q_0$, and $j_0$, respectively. The curves show the results obtained for aLIGO, the Einstein Telescope, and DECIGO. The vertical dotted lines indicate the reference catalogue sizes adopted for each interferometer: $N=10$, $100$, and $800$, respectively.}

\label{fig:sigma_N_all}
\end{figure*}

\subsection{Parameter Estimation} \label{sec: param_est}

In this work, we adopt a Bayesian framework~\cite{Trotta:2017wnx} to estimate the parameters $\{\theta_i\} = \{H_0, q_0, j_0\}$ given the model $\mathcal{M}$, represented by the cosmographic expansion in Eq.~\eqref{dl_cosmography}, truncated at third order. In this context, the joint posterior probability distribution of the parameters $\theta_i$ is given by

\begin{equation}
P(\{\theta_i\}|d, \mathcal{M}) = \frac{P(d|\{\theta_i\}, \mathcal{M})\, P(\{\theta_i\}|\mathcal{M})}{P(d)},
\end{equation}
where the likelihood $P(d|\{\theta_i\}, \mathcal{M})$ is defined as
\begin{equation}
P(d|\{\theta_i\}, \mathcal{M}) = \exp \left[ -\frac{1}{2} \sum_{i=1}^{N} 
\frac{|d_i - f_i(\{\theta_i\}, z_i)|^2}{\sigma_i^2} \right],
\end{equation}
with $N$ being the number of detections, $d_i$ the observed data with uncertainty $\sigma_i$, and $f_i({\theta_i},z_i)$ the model prediction. The simulated catalogues are generated assuming the fiducial flat $\Lambda$CDM cosmology introduced previously, with $H_0=67.66$ $\mathrm{km\,s^{-1}\,Mpc^{-1}}$ and $\Omega_m=0.311$. The priors on $H_0$, $q_0$, and $j_0$ are assumed to be uniform within the ranges $(54,81)$ $\mathrm{km\,s^{-1}\,Mpc^{-1}}$, $(-1,0)$, and $(-1,3)$, respectively, chosen to encompass current observational constraints while remaining sufficiently broad to avoid informative priors. We note that the uncertainty in the redshift is neglected, consistently with the assumption of an electromagnetic counterpart to the gravitational wave signal, which allows for the unambiguous identification of the host galaxy.

\section{Results} \label{sec: results}

In this section, we apply the previously described methodology to investigate how different observational regimes of gravitational-wave detectors affect the constraints on the cosmographic parameters. We consider two approaches based on the number of detections ($N$) and the signal-to-noise ratio (SNR). The analysis as a function of $N$ constitutes our main investigation, whereas the SNR-based study is presented as a complementary assessment. Together, they allow us to examine how the catalogue size and the properties of the detected signals affect the precision of $H_0$, $q_0$ and $j_0$.

In the primary analysis, we construct mock catalogues containing different numbers of detected events and evaluate how the cosmographic constraints evolve as $N$ increases. This approach directly quantifies the statistical gain provided by progressively larger samples and enables comparisons among the interferometers at common values of $N$.\\
\indent For the complementary SNR analysis, the mock catalogues are constructed separately within predefined SNR intervals, so that all events in a given sample have signal-to-noise ratios inside the corresponding bin. The number of detections is fixed at $N=100$ in every interval and for all interferometers, thereby removing variations caused solely by different sample sizes. This controlled setup allows us to examine the behavior of the constraints across different SNR regimes while retaining the intrinsic correlation between SNR and redshift.\\
\indent For each bin in both analysis, we performed 50 independent realizations and estimated the average of the mean values and the average of the standard deviations across all realizations. The posterior distributions of $H_0$, $q_0$, and $j_0$ were obtained through a Markov Chain Monte Carlo (MCMC) approach using \texttt{emcee}~\cite{Foreman_Mackey_2013}. A comprehensive summary of the adopted values of $N$, the SNR intervals, and the MCMC-derived cosmographic results is presented in~\ref{appendix: N} and~\ref{appendix: SNR}. In addition, Figures~\ref{fig:sigma_N_all} and~\ref{fig:sigma_SNR_all} illustrate the behavior of the parameter uncertainties as functions of $N$ and SNR, respectively, for all interferometers analyzed.

\subsection{N Analysis}

In the first part of the analysis, we investigate how the cosmographic constraints evolve as the number of detected events increases. To enable controlled comparisons among the interferometers, all three detectors are explored over the common interval $10\leq N\leq100$, while ET and DECIGO are additionally compared up to $N=800$. The analysis is then extended to $N=2000$ for DECIGO to investigate its behavior in more optimistic detection scenarios. These extended catalogue sizes remain within the expected numbers of detectable events for the corresponding interferometers, as summarized in Table~\ref{tab:N_initial}, and therefore do not represent event counts beyond their observational capabilities. All mock catalogues are restricted to the same redshift interval, $0<z<0.7$. The reference catalogue sizes, namely $N_z=10$, $100$, and $800$ for aLIGO, ET, and DECIGO, respectively, are indicated by the vertical dotted lines in Fig.~\ref{fig:sigma_N_all}. Extending ET and DECIGO toward smaller catalogues provides forecasts of the attainable precision during the initial stages of observation, before their reference event numbers are accumulated. Comparisons among different interferometers are performed only at common values of $N$, whereas the extensions beyond the overlapping ranges are used exclusively to examine the statistical improvement of each detector.

The top panel of Fig.~\ref{fig:sigma_N_all} shows that the uncertainty in the Hubble constant, $\sigma_{H_0}$, decreases with the number of detections for all interferometers. For aLIGO, it decreases from approximately $2.18~\mathrm{km/s/Mpc}$ at $N=10$ to $0.76~\mathrm{km/s/Mpc}$ at $N=100$. For ET, the corresponding uncertainty decreases from approximately $2.50~\mathrm{km/s/Mpc}$ at $N=10$ to $0.34~\mathrm{km/s/Mpc}$ at $N=800$, whereas DECIGO decreases from approximately $1.81~\mathrm{km/s/Mpc}$ at $N=10$ to $0.19~\mathrm{km/s/Mpc}$ at $N=2000$. At $N=100$, the corresponding uncertainties are approximately $0.76$, $1.07$, and $0.81~\mathrm{km/s/Mpc}$ for aLIGO, ET, and DECIGO, respectively. At the common value $N=800$, ET and DECIGO achieve $\sigma_{H_0}\simeq0.34$ and $0.30~\mathrm{km/s/Mpc}$, respectively.

The middle panel of Fig.~\ref{fig:sigma_N_all} presents the constraints on the deceleration parameter. For aLIGO, $\sigma_{q_0}$ changes only modestly, from approximately $0.337$ at $N=10$ to $0.310$ at $N=100$. In contrast, ET improves from $\sigma_{q_0}\simeq0.254$ at $N=10$ to $0.074$ at $N=800$, while DECIGO decreases from approximately $0.249$ at $N=10$ to $0.034$ at $N=2000$. At $N=100$, the corresponding uncertainties are approximately $0.310$, $0.196$, and $0.144$ for aLIGO, ET, and DECIGO, respectively. At the common value $N=800$, ET and DECIGO yield uncertainties of approximately $0.074$ and $0.054$, respectively.

A similar behavior is observed for the jerk parameter in the bottom panel of Fig.~\ref{fig:sigma_N_all}. The aLIGO uncertainty remains nearly constant, with $\sigma_{j_0}\simeq1.36$ throughout the interval $10\leq N\leq100$. For ET, $\sigma_{j_0}$ decreases from approximately $1.21$ at $N=10$ to $0.44$ at $N=800$, whereas DECIGO decreases from approximately $1.21$ at $N=10$ to $0.19$ at $N=2000$. At $N=100$, the corresponding uncertainties are approximately $1.36$, $1.05$, and $0.76$ for aLIGO, ET, and DECIGO, respectively. At the common value $N=800$, ET and DECIGO yield uncertainties of approximately $0.44$ and $0.29$, respectively. The complete numerical results are reported in Tables~\ref{tab:LIGO_N}, \ref{tab:ET_N}, and~\ref{tab:DECIGO_N}.

\subsection{SNR Analysis}

\begin{figure}[h]
\centering
\includegraphics[width=0.45\textwidth]
{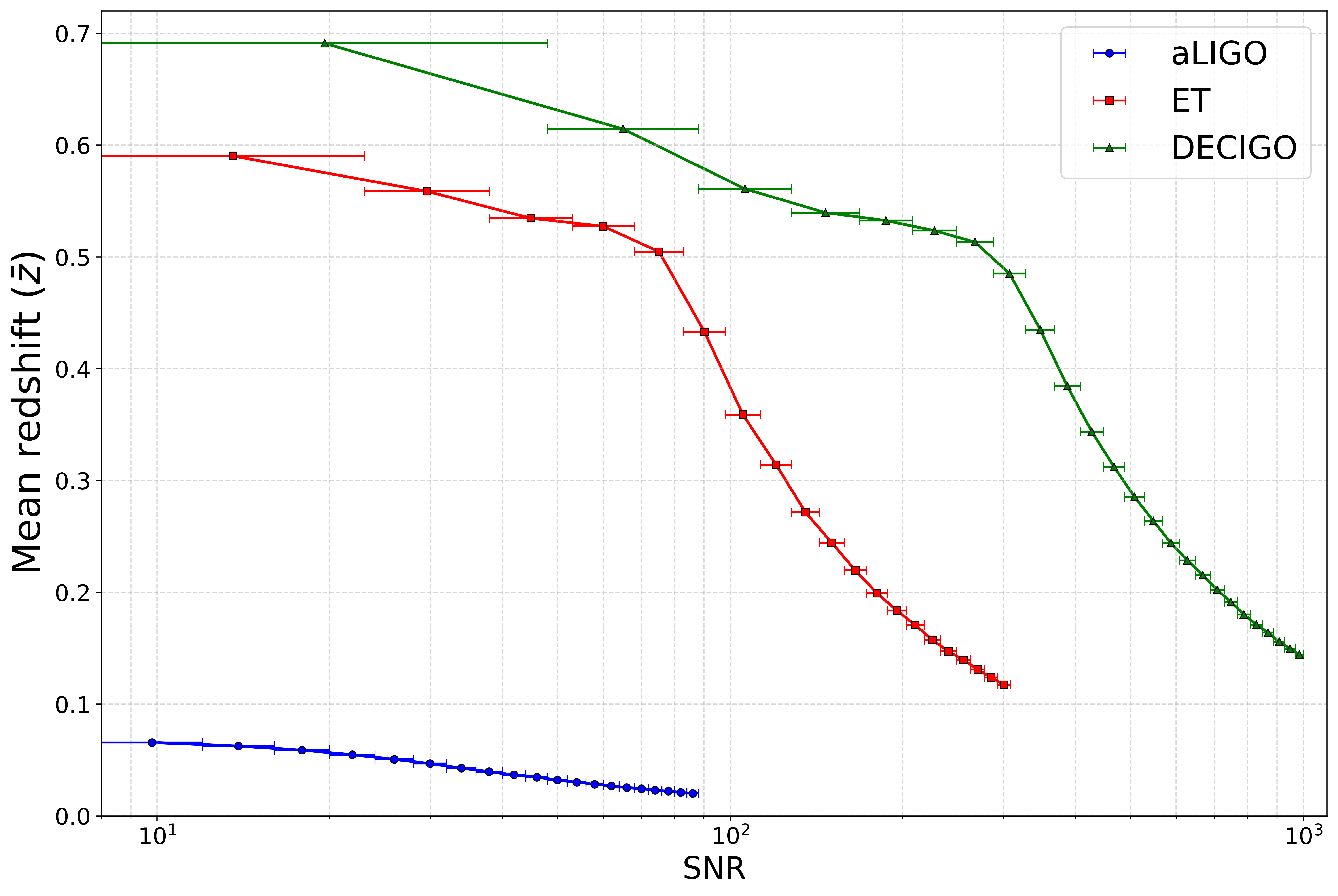}

\caption{Mean redshift of the mocks as a function of SNR for aLIGO (blue circles), ET (red squares), and DECIGO (green triangles). Each marker is positioned at the geometric center of its SNR bin, while the horizontal segments represent the complete detector-specific intervals. The results are averaged over 50 realizations with $N=100$ events per bin. The decrease in $\bar{z}$ with increasing SNR illustrates the correlation between these quantities.}

\label{fig:z_mean_SNR}
\end{figure}

As a complementary analysis, we construct mock catalogues within predefined, detector-specific SNR intervals. For each interval and each interferometer, source parameters are sampled until exactly $N=100$ accepted events are obtained. These fixed-size catalogues constitute controlled statistical samples and are not associated with a common observing time. Although the catalogue size is fixed, SNR and redshift cannot be varied independently because higher-SNR events preferentially originate at lower redshifts. We therefore calculate the mean redshift, $\bar{z}$, associated with each SNR interval and display this relation in Fig.~\ref{fig:z_mean_SNR}.

As shown in Fig.~\ref{fig:z_mean_SNR}, the mean redshift decreases systematically as the SNR increases for all three interferometers. The results below must therefore be interpreted as the joint evolution of the constraints with the SNR and redshift regimes at fixed $N$, rather than as a measurement of the causal effect of SNR alone. Moreover, because the detectors cover different SNR intervals and produce different redshift distributions, their curves should not be interpreted as a direct ranking of detector performance.

\begin{figure*}[t!]
\centering

\begin{subfigure}{\textwidth}
\centering
\includegraphics[width=0.61\textwidth]{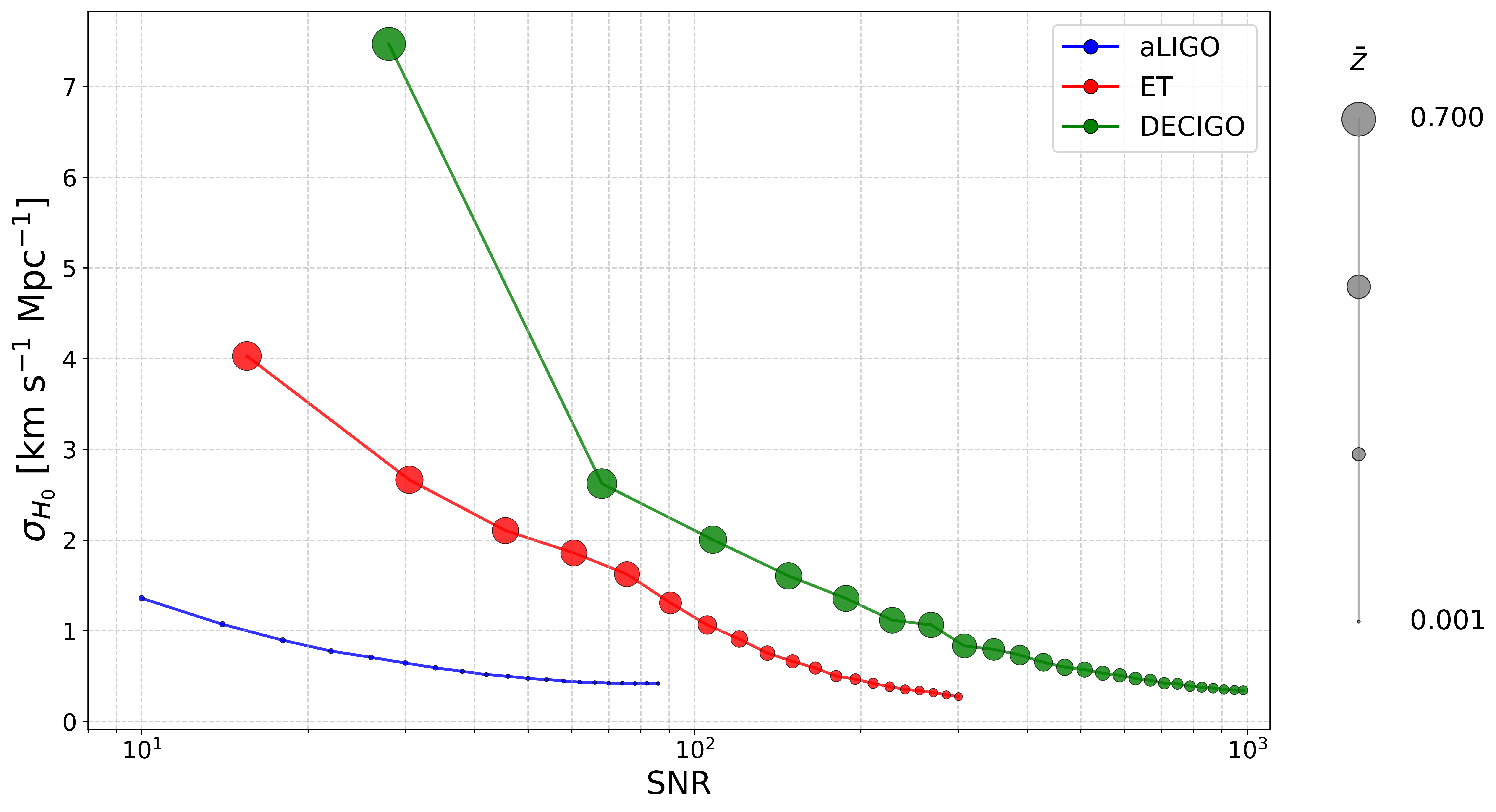}
%\caption{}
\label{fig:sigma_H0_SNR}
\end{subfigure}

%\vspace{1cm}

\begin{subfigure}{\textwidth}
\centering
\includegraphics[width=0.64\textwidth]{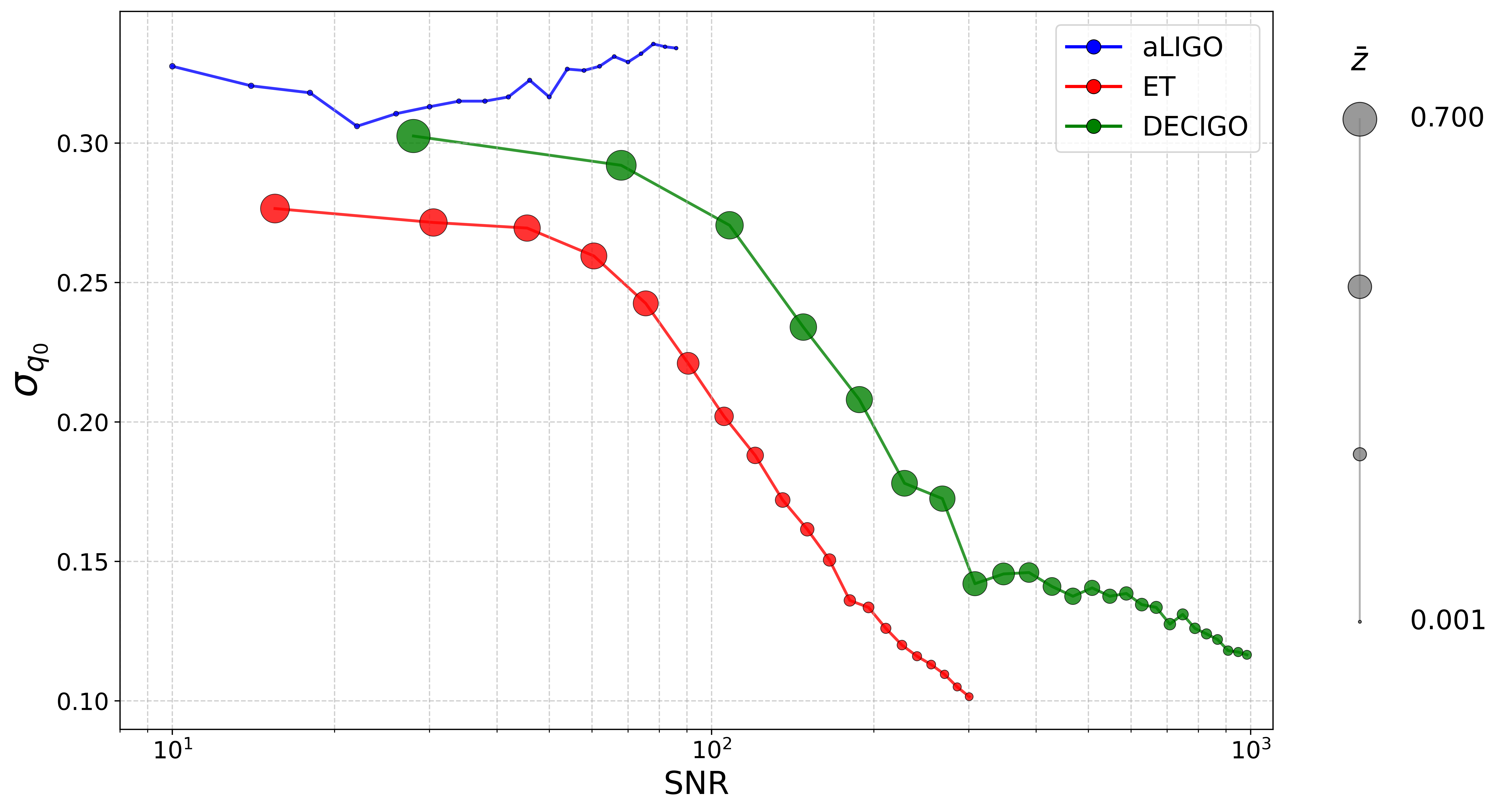}
%\caption{}
\label{fig:sigma_q0_SNR}
\end{subfigure}

%\vspace{1cm}

\begin{subfigure}{\textwidth}
\centering
\includegraphics[width=0.64\textwidth]{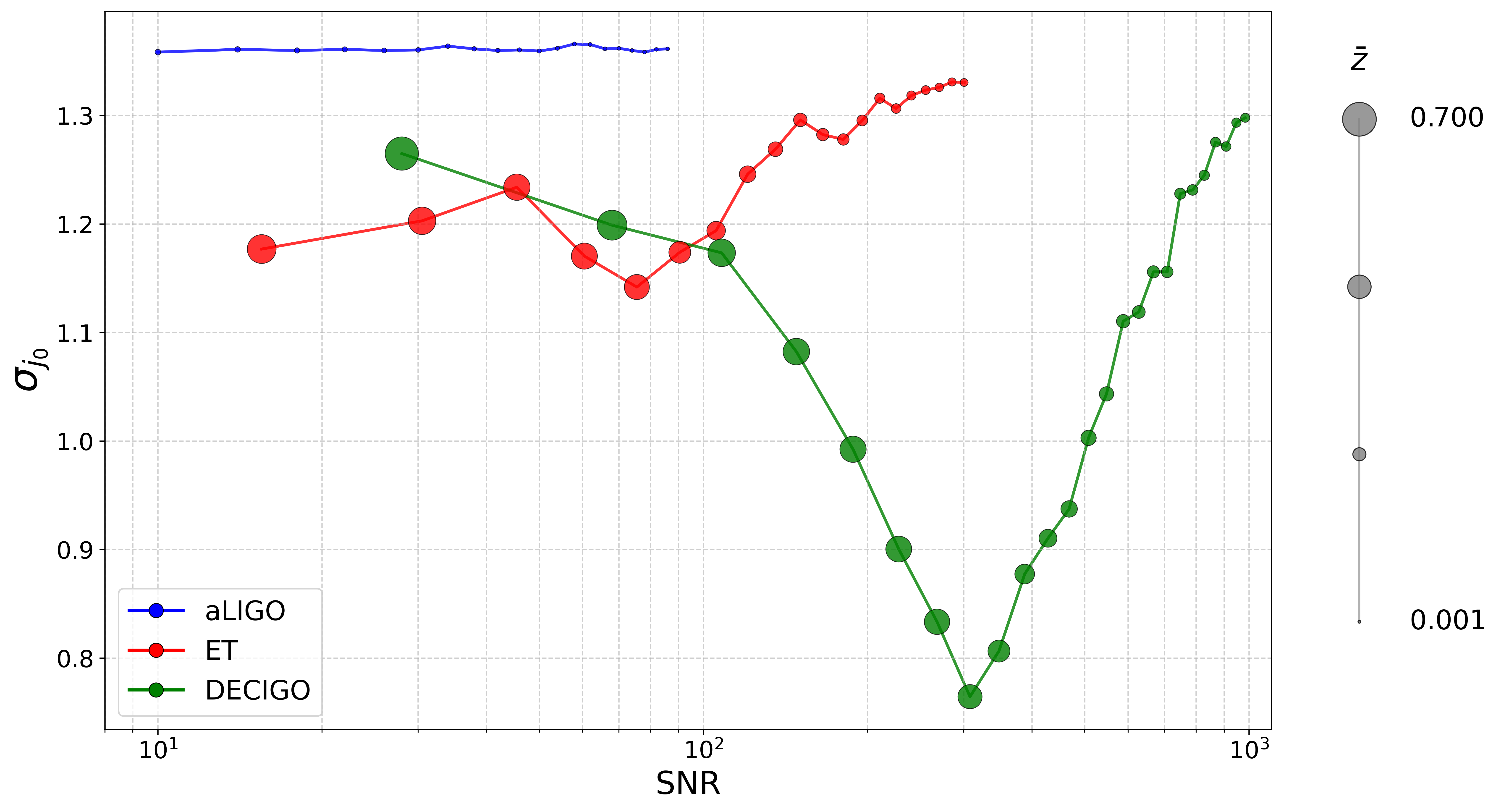}
%\caption{}
\label{fig:sigma_j0_SNR}
\end{subfigure}

\caption{
Uncertainties, $\sigma$, of the cosmographic parameters $H_0$, $q_0$, and $j_0$ as functions of the signal-to-noise ratio. The top, middle, and bottom panels correspond to $H_0$, $q_0$, and $j_0$, respectively. Each point is positioned at the central SNR value of its corresponding interval, while its size represents the associated mean redshift, $\bar{z}$. Larger circles therefore correspond to higher-redshift catalogues.}

\label{fig:sigma_SNR_all}
\end{figure*}

The top panel of Fig.~\ref{fig:sigma_SNR_all} shows that $\sigma_{H_0}$ decreases toward the higher-SNR and lower-redshift regime for all three interferometers. For aLIGO, it decreases from approximately $1.36$ $\mathrm{km\,s^{-1}\,Mpc^{-1}}$ in the interval $\mathrm{SNR}\in(8,12)$, where $\bar{z}=0.0657$, to $0.42$ $\mathrm{km\,s^{-1}\,Mpc^{-1}}$ in $\mathrm{SNR}\in(84,88)$, where $\bar{z}=0.0202$. For ET, the corresponding uncertainty changes from approximately $4.03$ $\mathrm{km\,s^{-1}\,Mpc^{-1}}$ in $\mathrm{SNR}\in(8,23)$, with $\bar{z}=0.5903$, to $0.28$ $\mathrm{km\,s^{-1}\,Mpc^{-1}}$ in $\mathrm{SNR}\in(293,308)$, with $\bar{z}=0.1174$. DECIGO exhibits the same overall trend, decreasing from approximately $7.47$ $\mathrm{km\,s^{-1}\,Mpc^{-1}}$ in $\mathrm{SNR}\in(8,48)$, where $\bar{z}=0.6911$, to $0.35$ $\mathrm{km\,s^{-1}\,Mpc^{-1}}$ in $\mathrm{SNR}\in(968,1000)$, where $\bar{z}=0.1443$.

The middle panel of Fig.~\ref{fig:sigma_SNR_all} presents the uncertainty in the deceleration parameter. For aLIGO, $\sigma_{q_0}$ remains approximately constant throughout the explored range, with a shallow minimum of about $0.306$ in $\mathrm{SNR}\in(20,24)$, where $\bar{z}=0.0548$, and values close to $0.33$ at the highest SNRs, where $\bar{z}$ decreases to $0.0202$. ET shows an overall decrease from approximately $0.277$ in $\mathrm{SNR}\in(8,23)$, with $\bar{z}=0.5903$, to $0.102$ in $\mathrm{SNR}\in(293,308)$, with $\bar{z}=0.1174$. Similarly, DECIGO decreases from approximately $0.303$ in $\mathrm{SNR}\in(8,48)$, where $\bar{z}=0.6911$, to $0.117$ in $\mathrm{SNR}\in(968,1000)$, where $\bar{z}=0.1443$, with small fluctuations along the curve.

The bottom panel of Fig.~\ref{fig:sigma_SNR_all} displays the uncertainty in the jerk parameter. The aLIGO curve remains nearly constant at $\sigma_{j_0}\simeq1.36$, showing no appreciable improvement across its SNR range. For ET, the uncertainty reaches a shallow minimum of approximately $1.14$ in $\mathrm{SNR}\in(68,83)$, where $\bar{z}=0.5048$, and subsequently increases to approximately $1.33$ in $\mathrm{SNR}\in(293,308)$, where $\bar{z}=0.1174$. DECIGO presents a more pronounced non-monotonic behavior: $\sigma_{j_0}$ decreases from approximately $1.27$ in $\mathrm{SNR}\in(8,48)$ to a minimum of $0.76$ in $\mathrm{SNR}\in(288,328)$, where $\bar{z}=0.4850$, and then increases to approximately $1.30$ in $\mathrm{SNR}\in(968,1000)$, where $\bar{z}=0.1443$. The complete numerical results are reported in Tables~\ref{tab:LIGO_SNR}, \ref{tab:ET_SNR}, and~\ref{tab:DECIGO_SNR}.

\section{Conclusions} \label{sec: conclusion}

In the present paper we investigated the capability of future gravitational wave observations with both current detectors, such as Advanced LIGO, as well as next-generation interferometers, including the Einstein Telescope and DECIGO, to constrain the parameters of the cosmographic expansion like the Hubble constant, deceleration and jerk parameters. It is well known that truncated cosmographic series are not expected to accurately describe the cosmic expansion beyond redshifts of order unity. As theoretically expected, including higher-order terms improves the agreement between the Taylor expansion and the fiducial $\Lambda$CDM model. We nevertheless adopt the third-order truncation as a conservative choice, supported by observational analyses showing that additional cosmographic parameters may increase degeneracies and statistical uncertainties. Once the truncation order is fixed, the comparison with the fiducial model shows that the relative deviation remains within approximately $1\%$ up to $z_{\max}=0.70$. The simulated catalogs assume the detection of electromagnetic counterparts, ensuring accurate redshift determination, with sources observed up to $z_{\max}$. We then examined the cosmographic constraints primarily as a function of the number of detections and, in a complementary analysis with $N=100$ fixed in every bin, as a function of SNR while explicitly accounting for its correlation with the mean redshift of the mock.

The accuracy achieved by gravitational-wave standard sirens has direct implications for the Hubble tension~\cite{DiValentino:2021izs, Hu:2023jqc, Capozziello:2024stm, Kraiselburd:2025gti}. For Advanced LIGO, the relative uncertainty in $H_0$ decreases from $3.2\%$ at $N=10$ to $1.1\%$ at $N=100$. At the common sample size $N=100$, aLIGO, ET, and DECIGO yield relative uncertainties of $1.1\%$, $1.6\%$, and $1.2\%$, respectively. At the common value $N=800$, ET and DECIGO reach $0.5\%$ and $0.4\%$, respectively, while the detector-specific extension of DECIGO to $N=2000$ yields $0.3\%$. The $1\%-2\%$ precision reached at $N=100$ could already allow standard sirens to discriminate meaningfully between the early and late-Universe determinations of $H_0$ if the value inferred from real observations lies sufficiently close to either of them, while the subpercent precision forecast for larger ET and DECIGO catalogues would provide even stronger discriminatory power. However, because the simulated catalogues adopt the fiducial value $H_0=67.66~\mathrm{km/s/Mpc}$, these forecasts establish the attainable precision but do not themselves favor either side of the Hubble tension.

The constraints on the deceleration parameter, $q_0$, quantify the sensitivity of standard sirens to the present acceleration rate. At the common sample size $N=100$, the relative uncertainties are $63.7\%$, $39.7\%$, and $27.9\%$ for aLIGO, ET, and DECIGO, respectively. At $N=800$, ET and DECIGO reach $14.3\%$ and $10.0\%$, while the DECIGO optimistic scenario with $N=2000$ reaches $6.3\%$. The recovered estimates remain consistent with the fiducial value and show that large future catalogues can measure the magnitude of $q_0$ with substantially greater precision.

The jerk parameter, $j_0$, provides a direct test of the expansion history beyond the deceleration term and is particularly useful for assessing departures from the standard flat $\Lambda$CDM prediction $j_0=1$. At $N=100$, the relative uncertainties are $134.0\%$, $135.9\%$, and $80.6\%$ for aLIGO, ET, and DECIGO, respectively. For the common sample size $N=800$, ET reaches a precision of $46.4\%$, whereas DECIGO reaches $28.9\%$. In the detector-specific DECIGO scenario with $N=2000$, the uncertainty decreases to $18.1\%$. As expected for catalogues generated from the fiducial $\Lambda$CDM model, all recovered constraints remain consistent with $j_0=1$ and do not indicate a deviation from the standard scenario. Instead, these forecasts quantify the level at which future standard-siren catalogues could test such departures, with the most optimistic DECIGO configuration reaching sensitivity at the few-tens-of-percent level. This prospect is particularly relevant in light of recent results from DESI~\cite{DESI:2025zgx, Rodrigues:2025tfg}, which have reported indications of possible departures from the cosmological constant. Future gravitational-wave observations may therefore provide a complementary and distance-ladder-independent avenue for investigating the dynamical nature of dark energy.

In the complementary SNR analysis, we fix the catalogue size at $N=100$ in every bin and investigate how the cosmographic precision varies across the correlated SNR and redshift regimes. For all detectors, higher-SNR bins are associated with lower mean redshifts, and the uncertainty in $H_0$ generally decreases toward this regime. For $q_0$, aLIGO remains approximately insensitive to these variations, whereas ET and DECIGO exhibit an overall improvement. The behavior of $j_0$ is distinct: the aLIGO uncertainty remains nearly constant, while ET and DECIGO achieve their tightest constraints at intermediate SNRs and subsequently lose precision toward the highest-SNR and lowest-redshift bins. This non-monotonic behavior reflects the interplay between SNR and redshift leverage for higher-order cosmographic parameters, in agreement with previous studies indicating that observations extending to $z\gtrsim0.4$ are important for probing these terms~\cite{deSouza:2021xtg}.

In summary, gravitational wave standard sirens provide a robust and versatile framework for cosmographic studies in the low-redshift Universe. Current-generation detectors can provide independent percent-level measurements of the Hubble constant, whereas large catalogues from next-generation and space-based interferometers extend this approach to higher-order cosmographic parameters. Under controlled comparisons at common sample sizes, future interferometers offer substantial improvements. As gravitational wave catalogues continue to grow in size and sensitivity, cosmography with standard sirens is poised to become a powerful complementary tool to traditional electromagnetic surveys in shaping our understanding of the expansion history of the Universe.

\section*{Acknowledgments}

JM acknowledges support from Coordenação de Aperfeiçoamento de Pessoal de Nível Superior (CAPES). RSG thanks the financial support from Fundação de Amparo à Pesquisa do Estado do Rio de Janeiro (FAPERJ) under Grant No. 260003/005977/2024 (APQ1). JSA is supported by Conselho Nacional de Desenvolvimento Científico e Tecnológico (CNPq) under Grant No. 307683/2022-2 and by FAPERJ under Grant No. 299312 (2023). This work made use of the National Observatory Data Center (CPDON).

\bibliographystyle{elsarticle-num} 
\bibliography{references}

\clearpage
\onecolumn
\raggedbottom
\appendix

\section{Numerical Results for the N-based Analysis}
\label{appendix: N}

\vspace{1em}

\begin{center}
\scriptsize
\setlength{\tabcolsep}{3pt}
\renewcommand{\arraystretch}{1.4}

\begin{adjustbox}{max width=\textwidth}
\begin{tabular}{|c|c|c|c|c|c|c|}
\hline
$N$
& {$H_0$ $[\mathrm{km\,s^{-1}\,Mpc^{-1}}]$}
& {$q_0$}
& {$j_0$}
& {$\sigma_{H_0}/H_0$ (\%)}
& {$\sigma_{q_0}/|q_0|$ (\%)}
& {$\sigma_{j_0}/j_0$ (\%)} \\
\hline

10  & $68.17^{+2.24}_{-2.11}$ & $-0.507^{+0.341}_{-0.333}$ & $0.998^{+1.359}_{-1.361}$ & 3.2 & 66.5 & 136.3 \\ \hline
20  & $67.78^{+1.55}_{-1.49}$ & $-0.506^{+0.335}_{-0.332}$ & $0.992^{+1.368}_{-1.353}$ & 2.2 & 65.9 & 137.1 \\ \hline
30  & $67.77^{+1.25}_{-1.21}$ & $-0.508^{+0.336}_{-0.330}$ & $0.985^{+1.378}_{-1.353}$ & 1.8 & 65.6 & 138.6 \\ \hline
40  & $67.60^{+1.10}_{-1.06}$ & $-0.516^{+0.336}_{-0.319}$ & $0.991^{+1.364}_{-1.350}$ & 1.6 & 63.5 & 136.9 \\ \hline
50  & $67.54^{+0.99}_{-0.97}$ & $-0.503^{+0.330}_{-0.326}$ & $0.993^{+1.367}_{-1.363}$ & 1.5 & 65.2 & 137.5 \\ \hline
60  & $67.85^{+0.95}_{-0.92}$ & $-0.497^{+0.328}_{-0.328}$ & $1.013^{+1.359}_{-1.370}$ & 1.4 & 66.0 & 134.7 \\ \hline
70  & $67.71^{+0.88}_{-0.85}$ & $-0.478^{+0.314}_{-0.327}$ & $1.022^{+1.354}_{-1.378}$ & 1.3 & 67.1 & 133.7 \\ \hline
80  & $67.78^{+0.82}_{-0.81}$ & $-0.505^{+0.316}_{-0.313}$ & $1.005^{+1.361}_{-1.354}$ & 1.2 & 62.3 & 135.1 \\ \hline
90  & $67.66^{+0.79}_{-0.78}$ & $-0.506^{+0.319}_{-0.311}$ & $1.007^{+1.354}_{-1.366}$ & 1.2 & 62.3 & 135.1 \\ \hline
100 & $67.65^{+0.77}_{-0.75}$ & $-0.487^{+0.309}_{-0.311}$ & $1.015^{+1.357}_{-1.364}$ & 1.1 & 63.7 & 134.0 \\ \hline
\end{tabular}
\end{adjustbox}

\captionof{table}{Cosmographic parameter estimates for aLIGO as a function of the number of detected events.}
\label{tab:LIGO_N}
\end{center}

% =========================================================
% Einstein Telescope
% =========================================================
\renewcommand{\arraystretch}{1.4}
\begin{table}[H]
\centering
\scriptsize
\setlength{\tabcolsep}{3pt}
\begin{adjustbox}{max width=\textwidth}
\begin{tabular}{|c|c|c|c|c|c|c|}
\hline
$N$
& {$H_0$ $[\mathrm{km\,s^{-1}\,Mpc^{-1}}]$}
& {$q_0$}
& {$j_0$}
& {$\sigma_{H_0}/H_0$ (\%)}
& {$\sigma_{q_0}/|q_0|$ (\%)}
& {$\sigma_{j_0}/j_0$ (\%)} \\
\hline
10  & $67.37^{+2.57}_{-2.42}$ & $-0.483^{+0.264}_{-0.243}$ & $0.893^{+1.296}_{-1.127}$ & 3.7 & 52.5 & 135.7 \\ \hline
20  & $67.28^{+2.01}_{-1.94}$ & $-0.489^{+0.258}_{-0.239}$ & $0.847^{+1.338}_{-1.129}$ & 2.9 & 50.8 & 145.6 \\ \hline
30  & $67.24^{+1.63}_{-1.62}$ & $-0.485^{+0.248}_{-0.230}$ & $0.823^{+1.325}_{-1.111}$ & 2.4 & 49.3 & 148.0 \\ \hline
40  & $67.52^{+1.47}_{-1.47}$ & $-0.494^{+0.235}_{-0.228}$ & $0.739^{+1.334}_{-1.080}$ & 2.2 & 46.9 & 163.3 \\ \hline
50  & $67.39^{+1.21}_{-1.26}$ & $-0.508^{+0.223}_{-0.201}$ & $0.944^{+1.208}_{-1.078}$ & 1.8 & 41.7 & 121.1 \\ \hline
60  & $67.47^{+1.21}_{-1.25}$ & $-0.489^{+0.218}_{-0.202}$ & $0.784^{+1.184}_{-1.036}$ & 1.8 & 42.9 & 141.6 \\ \hline
70  & $67.58^{+1.09}_{-1.15}$ & $-0.508^{+0.217}_{-0.197}$ & $0.868^{+1.204}_{-1.069}$ & 1.7 & 40.7 & 130.9 \\ \hline
80  & $67.58^{+1.09}_{-1.12}$ & $-0.500^{+0.206}_{-0.197}$ & $0.809^{+1.177}_{-1.009}$ & 1.6 & 40.3 & 135.1 \\ \hline
90  & $67.60^{+1.03}_{-1.07}$ & $-0.523^{+0.202}_{-0.183}$ & $0.994^{+1.107}_{-1.007}$ & 1.6 & 36.8 & 106.3 \\ \hline
100 & $67.57^{+1.05}_{-1.09}$ & $-0.494^{+0.200}_{-0.192}$ & $0.774^{+1.137}_{-0.967}$ & 1.6 & 39.7 & 135.9 \\ \hline
180 & $67.66^{+0.73}_{-0.80}$ & $-0.532^{+0.162}_{-0.146}$ & $1.035^{+0.910}_{-0.866}$ & 1.1 & 28.9 & 85.8 \\ \hline
260 & $67.52^{+0.62}_{-0.62}$ & $-0.502^{+0.130}_{-0.129}$ & $0.881^{+0.776}_{-0.690}$ & 0.9 & 25.8 & 83.2 \\ \hline
340 & $67.70^{+0.55}_{-0.57}$ & $-0.549^{+0.120}_{-0.113}$ & $1.156^{+0.711}_{-0.675}$ & 0.8 & 21.2 & 59.9 \\ \hline
420 & $67.67^{+0.44}_{-0.46}$ & $-0.536^{+0.100}_{-0.096}$ & $1.033^{+0.610}_{-0.571}$ & 0.7 & 18.3 & 57.2 \\ \hline
500 & $67.61^{+0.43}_{-0.43}$ & $-0.520^{+0.094}_{-0.093}$ & $0.938^{+0.569}_{-0.525}$ & 0.6 & 18.0 & 58.3 \\ \hline
580 & $67.74^{+0.40}_{-0.40}$ & $-0.551^{+0.087}_{-0.087}$ & $1.109^{+0.551}_{-0.511}$ & 0.6 & 15.8 & 47.9 \\ \hline
660 & $67.70^{+0.36}_{-0.36}$ & $-0.544^{+0.079}_{-0.079}$ & $1.085^{+0.500}_{-0.463}$ & 0.5 & 14.5 & 44.4 \\ \hline
740 & $67.52^{+0.37}_{-0.37}$ & $-0.502^{+0.079}_{-0.080}$ & $0.834^{+0.478}_{-0.438}$ & 0.5 & 15.8 & 54.9 \\ \hline
800 & $67.58^{+0.34}_{-0.34}$ & $-0.519^{+0.074}_{-0.074}$ & $0.945^{+0.456}_{-0.421}$ & 0.5 & 14.3 & 46.4 \\ \hline
\end{tabular}
\end{adjustbox}
\caption{Cosmographic parameter estimates for the Einstein Telescope as a function of the number of detected events.}
\label{tab:ET_N}
\end{table}

% =========================================================
% DECIGO
% =========================================================
\renewcommand{\arraystretch}{1.4}
\begin{table}[H]
\centering
\scriptsize
\setlength{\tabcolsep}{3pt}
\begin{adjustbox}{max width=\textwidth}
\begin{tabular}{|c|c|c|c|c|c|c|}
\hline
$N$
& {$H_0$ $[\mathrm{km\,s^{-1}\,Mpc^{-1}}]$}
& {$q_0$}
& {$j_0$}
& {$\sigma_{H_0}/H_0$ (\%)}
& {$\sigma_{q_0}/|q_0|$ (\%)}
& {$\sigma_{j_0}/j_0$ (\%)} \\
\hline
10   & $67.52^{+1.78}_{-1.83}$ & $-0.494^{+0.256}_{-0.241}$ & $0.731^{+1.354}_{-1.070}$ & 2.7 & 50.3 & 165.8 \\ \hline
20   & $67.29^{+1.54}_{-1.50}$ & $-0.462^{+0.222}_{-0.215}$ & $0.687^{+1.203}_{-0.974}$ & 2.3 & 47.3 & 158.4 \\ \hline
30   & $67.35^{+1.33}_{-1.43}$ & $-0.483^{+0.235}_{-0.216}$ & $0.780^{+1.210}_{-1.013}$ & 2.0 & 46.7 & 142.5 \\ \hline
40   & $67.51^{+1.13}_{-1.20}$ & $-0.498^{+0.209}_{-0.195}$ & $0.826^{+1.110}_{-0.959}$ & 1.7 & 40.6 & 125.2 \\ \hline
50   & $67.42^{+1.04}_{-1.07}$ & $-0.495^{+0.185}_{-0.179}$ & $0.873^{+1.000}_{-0.860}$ & 1.6 & 36.8 & 106.5 \\ \hline
60   & $67.53^{+0.94}_{-1.00}$ & $-0.509^{+0.174}_{-0.162}$ & $0.934^{+0.918}_{-0.826}$ & 1.4 & 33.0 & 93.4 \\ \hline
70   & $67.53^{+0.93}_{-0.97}$ & $-0.505^{+0.172}_{-0.162}$ & $0.905^{+0.915}_{-0.821}$ & 1.4 & 33.1 & 95.9 \\ \hline
80   & $67.37^{+0.88}_{-0.90}$ & $-0.473^{+0.158}_{-0.156}$ & $0.725^{+0.856}_{-0.736}$ & 1.3 & 33.2 & 109.8 \\ \hline
90   & $67.53^{+0.82}_{-0.87}$ & $-0.517^{+0.155}_{-0.146}$ & $0.989^{+0.838}_{-0.758}$ & 1.3 & 29.1 & 80.7 \\ \hline
100  & $67.57^{+0.79}_{-0.83}$ & $-0.516^{+0.148}_{-0.140}$ & $0.949^{+0.801}_{-0.728}$ & 1.2 & 27.9 & 80.6 \\ \hline
180  & $67.76^{+0.62}_{-0.63}$ & $-0.549^{+0.112}_{-0.110}$ & $1.123^{+0.643}_{-0.587}$ & 0.9 & 20.2 & 54.8 \\ \hline
260  & $67.71^{+0.55}_{-0.55}$ & $-0.541^{+0.098}_{-0.097}$ & $1.066^{+0.559}_{-0.508}$ & 0.8 & 18.0 & 50.0 \\ \hline
340  & $67.73^{+0.47}_{-0.47}$ & $-0.551^{+0.084}_{-0.084}$ & $1.132^{+0.488}_{-0.446}$ & 0.7 & 15.2 & 41.3 \\ \hline
420  & $67.64^{+0.42}_{-0.42}$ & $-0.526^{+0.075}_{-0.075}$ & $0.979^{+0.421}_{-0.390}$ & 0.6 & 14.3 & 41.4 \\ \hline
500  & $67.66^{+0.39}_{-0.39}$ & $-0.536^{+0.070}_{-0.070}$ & $1.032^{+0.401}_{-0.368}$ & 0.6 & 13.1 & 37.3 \\ \hline
580  & $67.70^{+0.36}_{-0.36}$ & $-0.543^{+0.065}_{-0.065}$ & $1.064^{+0.373}_{-0.345}$ & 0.5 & 12.0 & 33.7 \\ \hline
660  & $67.70^{+0.34}_{-0.33}$ & $-0.535^{+0.060}_{-0.060}$ & $1.017^{+0.341}_{-0.317}$ & 0.5 & 11.2 & 32.4 \\ \hline
740  & $67.69^{+0.31}_{-0.31}$ & $-0.540^{+0.056}_{-0.057}$ & $1.058^{+0.323}_{-0.301}$ & 0.5 & 10.5 & 29.5 \\ \hline
800  & $67.68^{+0.30}_{-0.29}$ & $-0.535^{+0.053}_{-0.054}$ & $1.016^{+0.304}_{-0.284}$ & 0.4 & 10.0 & 28.9 \\ \hline
933  & $67.62^{+0.28}_{-0.28}$ & $-0.528^{+0.050}_{-0.050}$ & $0.985^{+0.282}_{-0.264}$ & 0.4 & 9.5 & 27.7 \\ \hline
1067 & $67.69^{+0.26}_{-0.26}$ & $-0.539^{+0.047}_{-0.047}$ & $1.032^{+0.265}_{-0.252}$ & 0.4 & 8.7 & 25.0 \\ \hline
1200 & $67.65^{+0.25}_{-0.24}$ & $-0.529^{+0.044}_{-0.044}$ & $0.974^{+0.246}_{-0.234}$ & 0.4 & 8.3 & 24.6 \\ \hline
1333 & $67.67^{+0.23}_{-0.23}$ & $-0.534^{+0.042}_{-0.042}$ & $1.003^{+0.237}_{-0.225}$ & 0.3 & 7.9 & 23.0 \\ \hline
1467 & $67.68^{+0.22}_{-0.22}$ & $-0.534^{+0.040}_{-0.040}$ & $1.005^{+0.225}_{-0.213}$ & 0.3 & 7.5 & 21.8 \\ \hline
1600 & $67.66^{+0.21}_{-0.21}$ & $-0.531^{+0.038}_{-0.038}$ & $0.986^{+0.213}_{-0.204}$ & 0.3 & 7.2 & 21.1 \\ \hline
1733 & $67.67^{+0.20}_{-0.20}$ & $-0.533^{+0.036}_{-0.037}$ & $1.005^{+0.205}_{-0.197}$ & 0.3 & 6.8 & 20.0 \\ \hline
1867 & $67.67^{+0.20}_{-0.20}$ & $-0.534^{+0.035}_{-0.035}$ & $1.002^{+0.197}_{-0.190}$ & 0.3 & 6.6 & 19.3 \\ \hline
2000 & $67.68^{+0.19}_{-0.19}$ & $-0.539^{+0.034}_{-0.034}$ & $1.040^{+0.193}_{-0.184}$ & 0.3 & 6.3 & 18.1 \\ \hline
\end{tabular}
\end{adjustbox}
\caption{Cosmographic parameter estimates for DECIGO as a function of the number of detected events.}
\label{tab:DECIGO_N}
\end{table}

\clearpage

\section{Numerical Results for the SNR-based Analysis} \label{appendix: SNR}

% =========================================================
% aLIGO
% =========================================================
\begin{table}[H]
\centering
\scriptsize
\setlength{\tabcolsep}{3pt}
\renewcommand{\arraystretch}{1.4}
\begin{adjustbox}{max width=\textwidth}
\begin{tabular}{|c|c|c|c|c|c|c|c|}
\hline
SNR interval
& $\bar{z}$
& $H_0$ $[\mathrm{km\,s^{-1}\,Mpc^{-1}}]$
& $q_0$
& $j_0$
& $\sigma_{H_0}/H_0$ (\%)
& $\sigma_{q_0}/|q_0|$ (\%)
& $\sigma_{j_0}/j_0$ (\%) \\
\hline
(8,12)  & 0.0657 & $67.60^{+1.38}_{-1.34}$ & $-0.516^{+0.334}_{-0.321}$ & $0.999^{+1.361}_{-1.356}$ & 2.0 & 63.5 & 136.0 \\ \hline
(12,16) & 0.0626 & $67.72^{+1.09}_{-1.05}$ & $-0.488^{+0.317}_{-0.324}$ & $0.998^{+1.364}_{-1.358}$ & 1.6 & 65.7 & 136.4 \\ \hline
(16,20) & 0.0589 & $67.48^{+0.91}_{-0.89}$ & $-0.514^{+0.324}_{-0.312}$ & $1.003^{+1.357}_{-1.363}$ & 1.3 & 61.9 & 135.6 \\ \hline
(20,24) & 0.0548 & $67.55^{+0.78}_{-0.77}$ & $-0.539^{+0.320}_{-0.292}$ & $0.967^{+1.378}_{-1.344}$ & 1.1 & 56.8 & 140.7 \\ \hline
(24,28) & 0.0507 & $67.70^{+0.71}_{-0.70}$ & $-0.516^{+0.317}_{-0.304}$ & $1.007^{+1.358}_{-1.362}$ & 1.0 & 60.2 & 135.1 \\ \hline
(28,32) & 0.0469 & $67.74^{+0.65}_{-0.64}$ & $-0.515^{+0.320}_{-0.306}$ & $0.995^{+1.362}_{-1.359}$ & 1.0 & 60.8 & 136.7 \\ \hline
(32,36) & 0.0429 & $67.51^{+0.59}_{-0.59}$ & $-0.509^{+0.319}_{-0.311}$ & $1.009^{+1.358}_{-1.370}$ & 0.9 & 61.9 & 135.2 \\ \hline
(36,40) & 0.0395 & $67.55^{+0.56}_{-0.55}$ & $-0.489^{+0.311}_{-0.319}$ & $1.008^{+1.359}_{-1.364}$ & 0.8 & 64.4 & 135.1 \\ \hline
(40,44) & 0.0368 & $67.55^{+0.52}_{-0.51}$ & $-0.497^{+0.314}_{-0.319}$ & $1.001^{+1.361}_{-1.359}$ & 0.8 & 63.7 & 135.9 \\ \hline
(44,48) & 0.0346 & $67.59^{+0.50}_{-0.50}$ & $-0.520^{+0.330}_{-0.315}$ & $0.980^{+1.370}_{-1.351}$ & 0.7 & 62.0 & 138.8 \\ \hline
(48,52) & 0.0322 & $67.68^{+0.48}_{-0.47}$ & $-0.507^{+0.319}_{-0.314}$ & $1.000^{+1.359}_{-1.360}$ & 0.7 & 62.4 & 136.0 \\ \hline
(52,56) & 0.0302 & $67.61^{+0.47}_{-0.46}$ & $-0.486^{+0.321}_{-0.332}$ & $1.006^{+1.359}_{-1.365}$ & 0.7 & 67.2 & 135.4 \\ \hline
(56,60) & 0.0284 & $67.57^{+0.45}_{-0.45}$ & $-0.499^{+0.327}_{-0.325}$ & $0.995^{+1.370}_{-1.362}$ & 0.7 & 65.3 & 137.3 \\ \hline
(60,64) & 0.0269 & $67.62^{+0.43}_{-0.44}$ & $-0.519^{+0.335}_{-0.320}$ & $1.002^{+1.364}_{-1.367}$ & 0.6 & 63.1 & 136.3 \\ \hline
(64,68) & 0.0254 & $67.67^{+0.43}_{-0.43}$ & $-0.497^{+0.330}_{-0.332}$ & $0.993^{+1.366}_{-1.357}$ & 0.6 & 66.6 & 137.1 \\ \hline
(68,72) & 0.0244 & $67.62^{+0.43}_{-0.42}$ & $-0.500^{+0.330}_{-0.328}$ & $1.008^{+1.356}_{-1.368}$ & 0.6 & 65.8 & 135.1 \\ \hline
(72,76) & 0.0230 & $67.68^{+0.43}_{-0.42}$ & $-0.488^{+0.326}_{-0.338}$ & $1.012^{+1.354}_{-1.366}$ & 0.6 & 68.0 & 134.4 \\ \hline
(76,80) & 0.0222 & $67.68^{+0.42}_{-0.42}$ & $-0.519^{+0.344}_{-0.327}$ & $0.997^{+1.359}_{-1.358}$ & 0.6 & 64.6 & 136.3 \\ \hline
(80,84) & 0.0211 & $67.73^{+0.42}_{-0.42}$ & $-0.505^{+0.337}_{-0.332}$ & $0.994^{+1.362}_{-1.360}$ & 0.6 & 66.2 & 136.9 \\ \hline
(84,88) & 0.0202 & $67.64^{+0.42}_{-0.42}$ & $-0.507^{+0.337}_{-0.331}$ & $0.993^{+1.364}_{-1.359}$ & 0.6 & 65.9 & 137.1 \\ \hline
\end{tabular}
\end{adjustbox}
\caption{Cosmographic parameter estimates for the aLIGO interferometer as a function of the signal-to-noise ratio. The quantity $\bar{z}$ denotes the mean redshift associated with each SNR interval.}
\label{tab:LIGO_SNR}
\end{table}

% =========================================================
% Einstein Telescope
% =========================================================
\begin{table}[H]
\centering
\scriptsize
\setlength{\tabcolsep}{3pt}
\renewcommand{\arraystretch}{1.4}
\begin{adjustbox}{max width=\textwidth}
\begin{tabular}{|c|c|c|c|c|c|c|c|}
\hline
SNR interval
& $\bar{z}$
& $H_0$ $[\mathrm{km\,s^{-1}\,Mpc^{-1}}]$
& $q_0$
& $j_0$
& $\sigma_{H_0}/H_0$ (\%)
& $\sigma_{q_0}/|q_0|$ (\%)
& $\sigma_{j_0}/j_0$ (\%) \\
\hline
(8,23)    & 0.5903 & $67.66^{+4.20}_{-3.86}$ & $-0.510^{+0.294}_{-0.259}$ & $0.891^{+1.252}_{-1.102}$ & 6.0 & 54.2 & 132.1 \\ \hline
(23,38)   & 0.5587 & $67.62^{+2.72}_{-2.61}$ & $-0.495^{+0.282}_{-0.261}$ & $0.772^{+1.360}_{-1.046}$ & 3.9 & 54.8 & 155.8 \\ \hline
(38,53)   & 0.5347 & $67.54^{+2.08}_{-2.14}$ & $-0.494^{+0.281}_{-0.258}$ & $0.763^{+1.391}_{-1.077}$ & 3.1 & 54.6 & 161.7 \\ \hline
(53,68)   & 0.5274 & $67.30^{+1.87}_{-1.85}$ & $-0.446^{+0.257}_{-0.262}$ & $0.584^{+1.366}_{-0.975}$ & 2.8 & 58.2 & 200.4 \\ \hline
(68,83)   & 0.5048 & $67.17^{+1.59}_{-1.66}$ & $-0.470^{+0.252}_{-0.233}$ & $0.771^{+1.258}_{-1.026}$ & 2.4 & 51.6 & 148.1 \\ \hline
(83,98)   & 0.4329 & $67.35^{+1.26}_{-1.35}$ & $-0.491^{+0.231}_{-0.211}$ & $0.821^{+1.249}_{-1.099}$ & 1.9 & 45.0 & 143.0 \\ \hline
(98,113)  & 0.3589 & $67.40^{+1.05}_{-1.08}$ & $-0.480^{+0.202}_{-0.202}$ & $0.721^{+1.305}_{-1.083}$ & 1.6 & 42.1 & 165.6 \\ \hline
(113,128) & 0.3141 & $67.51^{+0.89}_{-0.93}$ & $-0.498^{+0.191}_{-0.185}$ & $0.769^{+1.335}_{-1.157}$ & 1.3 & 37.8 & 162.0 \\ \hline
(128,143) & 0.2717 & $67.53^{+0.74}_{-0.77}$ & $-0.509^{+0.176}_{-0.168}$ & $0.874^{+1.330}_{-1.208}$ & 1.1 & 33.8 & 145.2 \\ \hline
(143,158) & 0.2445 & $67.61^{+0.65}_{-0.68}$ & $-0.516^{+0.165}_{-0.158}$ & $0.876^{+1.352}_{-1.240}$ & 1.0 & 31.3 & 147.9 \\ \hline
(158,173) & 0.2196 & $67.56^{+0.58}_{-0.60}$ & $-0.512^{+0.154}_{-0.147}$ & $0.906^{+1.321}_{-1.244}$ & 0.9 & 29.4 & 141.6 \\ \hline
(173,188) & 0.1992 & $67.69^{+0.49}_{-0.51}$ & $-0.541^{+0.144}_{-0.128}$ & $1.082^{+1.246}_{-1.310}$ & 0.7 & 25.1 & 118.1 \\ \hline
(188,203) & 0.1838 & $67.58^{+0.47}_{-0.47}$ & $-0.510^{+0.134}_{-0.133}$ & $0.855^{+1.354}_{-1.237}$ & 0.7 & 26.2 & 151.5 \\ \hline
(203,218) & 0.1707 & $67.67^{+0.41}_{-0.43}$ & $-0.534^{+0.129}_{-0.123}$ & $0.975^{+1.329}_{-1.303}$ & 0.6 & 23.6 & 135.0 \\ \hline
(218,233) & 0.1575 & $67.66^{+0.38}_{-0.39}$ & $-0.536^{+0.125}_{-0.115}$ & $1.075^{+1.281}_{-1.332}$ & 0.6 & 22.4 & 121.5 \\ \hline
(233,248) & 0.1472 & $67.61^{+0.35}_{-0.36}$ & $-0.521^{+0.118}_{-0.114}$ & $0.936^{+1.344}_{-1.293}$ & 0.5 & 22.3 & 140.9 \\ \hline
(248,263) & 0.1395 & $67.68^{+0.34}_{-0.34}$ & $-0.531^{+0.115}_{-0.111}$ & $0.969^{+1.336}_{-1.311}$ & 0.5 & 21.3 & 136.6 \\ \hline
(263,278) & 0.1310 & $67.60^{+0.32}_{-0.32}$ & $-0.516^{+0.111}_{-0.108}$ & $0.930^{+1.356}_{-1.296}$ & 0.5 & 21.2 & 142.6 \\ \hline
(278,293) & 0.1239 & $67.66^{+0.29}_{-0.30}$ & $-0.530^{+0.107}_{-0.103}$ & $0.969^{+1.343}_{-1.319}$ & 0.4 & 19.8 & 137.4 \\ \hline
(293,308) & 0.1174 & $67.63^{+0.27}_{-0.28}$ & $-0.525^{+0.103}_{-0.100}$ & $0.994^{+1.334}_{-1.327}$ & 0.4 & 19.3 & 133.9 \\ \hline
\end{tabular}
\end{adjustbox}
\caption{Cosmographic parameter estimates for the Einstein Telescope as a function of the signal-to-noise ratio. The quantity $\bar{z}$ denotes the mean redshift associated with each SNR interval.}
\label{tab:ET_SNR}
\end{table}

% =========================================================
% DECIGO
% =========================================================
\begin{table}[H]
\centering
\scriptsize
\setlength{\tabcolsep}{3pt}
\renewcommand{\arraystretch}{1.4}
\begin{adjustbox}{max width=\textwidth}
\begin{tabular}{|c|c|c|c|c|c|c|c|}
\hline
SNR interval
& $\bar{z}$
& $H_0$ $[\mathrm{km\,s^{-1}\,Mpc^{-1}}]$
& $q_0$
& $j_0$
& $\sigma_{H_0}/H_0$ (\%)
& $\sigma_{q_0}/|q_0|$ (\%)
& $\sigma_{j_0}/j_0$ (\%) \\
\hline
(8,48)     & 0.6911 & $68.17^{+7.60}_{-7.34}$ & $-0.529^{+0.323}_{-0.282}$ & $0.978^{+1.247}_{-1.283}$ & 11.0 & 57.2 & 129.3 \\ \hline
(48,88)    & 0.6145 & $67.34^{+2.61}_{-2.63}$ & $-0.484^{+0.307}_{-0.277}$ & $0.795^{+1.383}_{-1.015}$ & 3.9 & 60.3 & 150.8 \\ \hline
(88,128)   & 0.5609 & $67.29^{+1.95}_{-2.06}$ & $-0.479^{+0.282}_{-0.259}$ & $0.782^{+1.321}_{-1.026}$ & 3.0 & 56.5 & 150.1 \\ \hline
(128,168)  & 0.5396 & $67.58^{+1.53}_{-1.68}$ & $-0.506^{+0.247}_{-0.221}$ & $0.887^{+1.166}_{-0.999}$ & 2.4 & 46.2 & 122.0 \\ \hline
(168,208)  & 0.5325 & $67.39^{+1.29}_{-1.42}$ & $-0.503^{+0.219}_{-0.197}$ & $0.930^{+1.055}_{-0.930}$ & 2.0 & 41.4 & 106.7 \\ \hline
(208,248)  & 0.5235 & $67.66^{+1.06}_{-1.17}$ & $-0.535^{+0.188}_{-0.168}$ & $1.079^{+0.932}_{-0.869}$ & 1.6 & 33.3 & 83.5 \\ \hline
(248,288)  & 0.5133 & $67.34^{+1.05}_{-1.08}$ & $-0.480^{+0.175}_{-0.170}$ & $0.816^{+0.898}_{-0.769}$ & 1.6 & 35.9 & 102.1 \\ \hline
(288,328)  & 0.4850 & $67.75^{+0.80}_{-0.86}$ & $-0.545^{+0.147}_{-0.137}$ & $1.101^{+0.791}_{-0.738}$ & 1.2 & 26.1 & 69.4 \\ \hline
(328,368)  & 0.4351 & $67.52^{+0.78}_{-0.82}$ & $-0.517^{+0.150}_{-0.141}$ & $0.985^{+0.842}_{-0.771}$ & 1.2 & 28.1 & 81.9 \\ \hline
(368,408)  & 0.3844 & $67.61^{+0.71}_{-0.75}$ & $-0.524^{+0.150}_{-0.142}$ & $0.971^{+0.910}_{-0.845}$ & 1.1 & 27.9 & 90.4 \\ \hline
(408,448)  & 0.3438 & $67.67^{+0.63}_{-0.68}$ & $-0.534^{+0.147}_{-0.135}$ & $1.045^{+0.925}_{-0.896}$ & 1.0 & 26.4 & 87.1 \\ \hline
(448,488)  & 0.3121 & $67.61^{+0.58}_{-0.61}$ & $-0.520^{+0.141}_{-0.134}$ & $0.967^{+0.958}_{-0.917}$ & 0.9 & 26.4 & 96.9 \\ \hline
(488,528)  & 0.2853 & $67.55^{+0.57}_{-0.58}$ & $-0.504^{+0.141}_{-0.140}$ & $0.828^{+1.051}_{-0.955}$ & 0.9 & 27.9 & 121.1 \\ \hline
(528,568)  & 0.2638 & $67.60^{+0.53}_{-0.54}$ & $-0.516^{+0.138}_{-0.137}$ & $0.875^{+1.090}_{-0.997}$ & 0.8 & 26.6 & 119.3 \\ \hline
(568,608)  & 0.2440 & $67.68^{+0.49}_{-0.53}$ & $-0.539^{+0.145}_{-0.132}$ & $1.065^{+1.110}_{-1.111}$ & 0.8 & 25.7 & 104.3 \\ \hline
(608,648)  & 0.2286 & $67.64^{+0.47}_{-0.48}$ & $-0.519^{+0.135}_{-0.134}$ & $0.859^{+1.166}_{-1.072}$ & 0.7 & 25.9 & 130.3 \\ \hline
(648,688)  & 0.2153 & $67.60^{+0.45}_{-0.47}$ & $-0.519^{+0.136}_{-0.131}$ & $0.919^{+1.182}_{-1.130}$ & 0.7 & 25.7 & 125.8 \\ \hline
(688,728)  & 0.2023 & $67.64^{+0.41}_{-0.43}$ & $-0.530^{+0.131}_{-0.124}$ & $1.005^{+1.163}_{-1.149}$ & 0.6 & 24.1 & 115.0 \\ \hline
(728,768)  & 0.1913 & $67.61^{+0.41}_{-0.42}$ & $-0.515^{+0.131}_{-0.131}$ & $0.871^{+1.281}_{-1.175}$ & 0.6 & 25.4 & 141.0 \\ \hline
(768,808)  & 0.1802 & $67.62^{+0.38}_{-0.40}$ & $-0.526^{+0.130}_{-0.122}$ & $0.985^{+1.235}_{-1.228}$ & 0.6 & 24.0 & 125.0 \\ \hline
(808,848)  & 0.1712 & $67.61^{+0.38}_{-0.38}$ & $-0.517^{+0.124}_{-0.124}$ & $0.885^{+1.290}_{-1.200}$ & 0.6 & 24.0 & 140.7 \\ \hline
(848,888)  & 0.1641 & $67.64^{+0.36}_{-0.37}$ & $-0.530^{+0.125}_{-0.119}$ & $0.983^{+1.287}_{-1.264}$ & 0.5 & 23.0 & 129.8 \\ \hline
(888,928)  & 0.1558 & $67.70^{+0.35}_{-0.36}$ & $-0.537^{+0.120}_{-0.116}$ & $0.979^{+1.287}_{-1.256}$ & 0.5 & 22.0 & 129.9 \\ \hline
(928,968)  & 0.1497 & $67.63^{+0.34}_{-0.35}$ & $-0.521^{+0.118}_{-0.117}$ & $0.909^{+1.332}_{-1.255}$ & 0.5 & 22.6 & 142.3 \\ \hline
(968,1000) & 0.1443 & $67.59^{+0.34}_{-0.35}$ & $-0.515^{+0.118}_{-0.115}$ & $0.943^{+1.324}_{-1.272}$ & 0.5 & 22.6 & 137.6 \\ \hline
\end{tabular}
\end{adjustbox}
\caption{Cosmographic parameter estimates for DECIGO as a function of the signal-to-noise ratio. The quantity $\bar{z}$ denotes the mean redshift associated with each SNR interval.}
\label{tab:DECIGO_SNR}
\end{table}

%%       \cite{<label>} ==> [1]

%%
%Example citation, See \cite{lamport94}.




%\begin{thebibliography}{00}

%% For numbered reference style

%\bibitem{lamport94}
  %Leslie Lamport,
  %\textit{\LaTeX: a document preparation system},
  %Addison Wesley, Massachusetts,
  %2nd edition,
  %1994.

%\end{thebibliography}
\end{document}